\documentclass[fleqn,usenatbib]{mnras}

\usepackage{newtxtext,newtxmath}

\usepackage[T1]{fontenc}

\DeclareRobustCommand{\VAN}[3]{#2}
\let\VANthebibliography\thebibliography
\def\thebibliography{\DeclareRobustCommand{\VAN}[3]{##3}\VANthebibliography}

\usepackage{graphicx}	
\usepackage{amsmath}	
\usepackage{mathtools}
\usepackage{lineno}

\title[Lensing of Fast Transient Event Rates]{The Effect of Gravitational Lensing on Fast Transient Event Rates}

\author[M. W. Sammons et al.]{
Mawson W. Sammons,$^{1}$
C. W. James,$^{1}$
C. M. Trott$^{1}$
and M. Walker$^{2}$
\\
$^{1}$International Centre for Radio Astronomy Research, Curtin University, Bentley, WA 6102, Australia\\
$^{2}$Manly Astrophysics, 15/41-42 East Esplanade, Manly, NSW 2095, Australia
}

\date{Accepted XXX. Received YYY; in original form ZZZ}

\pubyear{2021}

\begin{document}
\label{firstpage}
\pagerange{\pageref{firstpage}--\pageref{lastpage}}
\maketitle

\begin{abstract}
Fast cosmological transients such as fast radio bursts (FRBs) and gamma-ray bursts (GRBs) represent a class of sources more compact than any other cosmological object. As such they are sensitive to significant magnification via gravitational lensing from a class of lenses which are not well-constrained by observations today. Low-mass primordial black holes are one such candidate which may constitute a significant fraction of the Universe's dark matter. Current observations only constrain their density in the nearby Universe, giving fast transients from cosmological distances the potential to form complementary constraints. Motivated by this, we calculate the effect that gravitational lensing from a cosmological distribution of compact objects would have on the observed rates of FRBs and GRBs. For static lensing geometries, we rule out the prospect that all FRBs are gravitationally lensed for a range of lens masses and show that lens masses greater than $10^{-5}M_\odot$ can be constrained with 8000 un-localised high fluence FRBs at 1.4GHz, as might be detected by the next generation of FRB-finding telescopes.
\end{abstract}


\section{Introduction}\label{sec:intro}

The observed number of sources as a function of flux (the `logN--logS' relation) is perhaps the most fundamental quantity to population studies across astronomy. Encoded within them are the properties of the source population: luminosity function, spectra and redshift distribution \citep{longair_interpretation_1966}. Constraining the population functions of fast radio bursts (FRBs) and gamma-ray bursts (GRBs) has been an area of particular interest in modern astronomy\citep{macquart_frb_2018,sun_extragalactic_2015}. These bursts are often produced in exotic systems or cataclysmic circumstances\citep{abbott_gravitational_2017, woosley_supernova_2006, platts_living_2019} and therefore represent an important tracer of rare systems.

Propagation effects such as extinction, absorption or scintillation \citep{frontera_prompt_2000, masui_dense_2015} can however serve to obfuscate the intrinsic behaviours of a transient source, and thus must be accounted for if we are to determine the population functions from the observations. One such propagation effect is gravitational lensing.

Gravitational lensing is important to consider because it can magnify source objects, significantly amplifying their observed flux, potentially resulting in erroneously inferred luminosities. Many examples of lenses magnifying, distorting or even multiply imaging individual sources have been recorded, including for quasars \citep{walsh_0957_1979}, GRBs \citep{paynter_evidence_2021} and supernovae (SNe) \citep{kelly_multiple_2015}. However, the influence of gravitational lensing on the source counts of a population is typically slight; the fraction of quasars undergoing strong lensing is expected to be only a few percent of quasars beyond redshift six \citep{pacucci_most_2019, yue_revisiting_2022} and for SNe \citep{porciani_gravitational_2000, jonsson_constraining_2010} the fraction of lensed bursts is constrained to be small.

The lack of observed gravitational lensing can be used to infer constraints on the population of lenses. \citet{zumalacarregui_limits_2018} placed strong constraints on the fraction of dark matter in primordial black holes (PBHs) using type Ia SNe. The authors model the probability of magnification convolved with the spread in supernova magnitudes and compare with the observed spread to constrain the population of lenses at cosmological distances. They find that the observations are inconsistent with a large population of lenses and therefore restrict the fraction of dark matter in PBHs to be less than 0.3 for PBH masses greater than $0.01M_\odot$. Strong constraints can be placed in the case of SNe Ia observations because of the narrow distribution of intrinsic SNe Ia energies (i.e.\ because SNe Ia are standard candles).

These results do not mean that lensing will be unimportant for FRB and GRB source counts. The smaller angular size of GRBs and FRBs compared to SNe Ia makes them sensitive to even lower mass lenses. The uncertainty surrounding the emission mechanism of most fast transients, however, makes it difficult to separate potential propagation effects from potential emission mechanism effects. For example, it is difficult to distinguish a highly magnified event from an intrinsically luminous burst. Thus, lensing searches have been restricted to searches for multiple source images, be it in the spatial or temporal domain \citep{munoz_lensing_2016, laha_lensing_2018, oguri_strong_2019, sammons_first_2020, paynter_evidence_2021, leung_constraining_2022, chimefrb_collaboration_high-time_2022, connor_stellar_2022}.

As we shall show, the luminosity of GRBs and FRBs are sensitive to lens masses much too small to produce resolvable multiple images: as small as $10^{-15}M_\odot$ and $10^{-5}M_\odot$ respectively, and there is little evidence to rule out the presence of a cosmological population of lenses on these low mass scales. Constraints on primordial black holes (PBHs) still allow for $100\%$ of dark matter to be comprised of PBHs in the asteroid to sub-lunar mass regime \citep[$10^{-15}M_\odot\leq M_L\geq 10^{-10}M_\odot$;][]{carr_primordial_2020}. The sub-lunar to sub-stellar ($10^{-10}M_\odot\leq M_L\geq 10^{-2}M_\odot$) regime is also of interest as it is only constrained for our own galaxy halo, with $\sim 10^{-5}M_\odot$ lenses potentially existing locally in appreciable density.

To account for lensing effects on the source counts of fast transients, and estimate how this may be used to constrain the number of PBHs, we create a generic model for differential rates of fast transients with fluence, $dR/dF$, in an inhomogeneous universe and compare it to its smooth universe counterpart, considering only the total magnification caused by inhomogeneity. 

This paper is structured as follows:
\S \ref{subsec:lensingBasics} $\&$ \S\ref{subsec:PDF} introduce the lensing theory,
\S \ref{subsec:smoothRates} $\&$ \ref{subsec:clumpyRates} introduce the differential event rates formalism,
\S \ref{subsec:numImplementation} discusses our numerical method and justifies our model's assumptions, in \S \ref{sec:IntrinsicParamVar} we characterise the response of our model to variation of the input parameters, \S \ref{sec:AllLensed} explores the possibility that all FRBs are highly magnified, \S \ref{sec:specificCases} contains explicit calculations of the changes to FRB and GRB event rates in universes of varying inhomogeneity and finally, we discuss the implications of these results in \S \ref{sec:discussion} and explore how many FRBs would be needed to place constraints on the PBH parameter space.

\section{Method}\label{sec:theory}
\subsection{Lensing Basics}\label{subsec:lensingBasics}

Gravitational lensing is a result of perturbations in the mass of a smooth universe deflecting the emission of background sources. By convention we define the fraction of matter in the Universe which may be considered as smoothly distributed as $\eta$. A completely homogeneous universe ($\eta=1$) will be devoid of any gravitational lensing, i.e. the flux from a source at a given redshift will be constant for every line of sight. Whereas, an inhomogeneous universe will have a fraction of its total energy density $(1-\eta)\Omega_M$ in lensing objects and a corresponding distribution of possible magnifications associated with a given redshift. 

In general lensing causes a rich variety of effects on source images and temporal profiles of transients \citep[for a detailed review of which we refer the reader to][]{schneider_gravitational_1992}. In this work we restrict ourselves to consideration only of the total magnification of a source by a lens, 
\begin{equation}
    \mu=\frac{F}{F_0},
\end{equation}
where $F$ is the sum of the fluence from all images and $F_0$ is the fluence observed from a source along an `empty beam'. The empty beam is defined as the path of propagation which lies far from all clumps of inhomogeneous matter. $D_\eta$ is the value of angular diameter distance $D_A$ along the empty beam and it represents the background value ($\mu=1$) of $D_A$ in a universe with a smooth matter fraction $\eta$. Following the method outlined by \cite{kayser_general_1997}, $D_\eta$ can be calculated numerically for a general choice of both cosmology and $\eta$ (see appendix \ref{app:Deta} for extended discussion).

A critical quantity of the magnification distribution is the mean source magnification at a given redshift,
\begin{equation}\label{eq:meanMagnification}
    \langle\mu\rangle=\frac{D_\eta^2(z)}{D_1^2(z)},
\end{equation}
where $D_1$ represents the typical angular diameter distance of a smooth universe ($\eta=1$). As $\eta$ is increased and the universe becomes homogeneous, $D_\eta$ tends towards $D_1$ and we recover the smooth universe behaviour of $\langle\mu\rangle$=1. The mean magnification determines the shape of the magnification probability density function (PDF) we apply from \cite{rauch_gravitational_1991}.

\subsection{Magnification Probability Density Function}\label{subsec:PDF}

To determine the effects of gravitational lensing on observed fast transient event rates, we require a functional form for the magnification PDF. In this work we make use of the analytical approximation detailed in \cite{rauch_gravitational_1991},
\begin{equation}\label{eq:PDF}
    p(\mu) = 2\sigma_{\text{eff}}\left[\frac{1-e^{-b(\mu-1)}}{\mu^2-1}\right]^{1.5},
\end{equation}
where parameters $\sigma$ and $b$ are chosen such that the PDF is normalised and has a mean magnification $\langle\mu\rangle$. The form of the PDF is derived empirically by fitting to simulations of lensing rather than being motivated physically. However, by doing so it implicitly accounts for multiple lensing and shear which are significant complexities to hurdle when deriving a more physical model \citep{schneider_light_1988}. 

As stated in \cite{rauch_gravitational_1991} this approximation is only valid for low mean magnifications and point sources. However, even for large mean magnifications where the lensing enters the complex regime associated with an intricate caustic network, the approximation by Rauch provides a simple way to capture broad behaviour of the magnification probability distributions found by numerical simulations \citep[e.g.][]{fleury_simple_2020}. Given the relative uncertainties associated with both FRB and GRB luminosity functions \citep[][]{james_fast_2021, banerjee_differential_2021} particularly FRBs as their progenitor/s remain unknown, we will make use of this simple empirical model as opposed to vastly more computationally intensive numerical simulations. Finally, we note that this model is only valid for a stationary universe. If the magnification of a source can change significantly over time due to its motion relative to the lens then the probability of a given magnification must be reconsidered under a different formalism. We assume that both lenses and sources are stationary relative to the observer for the remainder of this work.

\subsection{Rates in a Smooth Universe}\label{subsec:smoothRates}
The impulsive nature of fast transient events means that burst rates rather than source counts are the fundamental quantity to consider when characterising the population. Furthermore, the observable directly relevant to transient events is fluence rather than flux which is typical for continuous sources. We use the fluence--energy relation outlined in \cite{macquart_frb_2018}.

The observed rate of a transient population is primarily governed by the intrinsic event rate energy function which depends on the redshift of the burst as well as its spectral energy and frequency in the emission frame $\Theta_E(z, E_{\nu_e}, \nu_e)$. This function yields the event rate per spectral energy per co-moving volume. Assuming that the redshift, spectral energy and emission frequency of a burst are independent, it can be separated into the population functions describing each dimension,
\begin{equation}
    \Theta_E(z, E_{\nu_e}, \nu_e)= \theta_z(z)\;\theta_{E}(E_{\nu_e}, E_{\nu_e,\text{max}},\gamma)\;\theta_{\nu_e}(\nu_e,\alpha),
\end{equation}
allowing us to motivate the form of the $\theta_x$ functions separately, depending on which transient we are considering. Above we have labelled each of the functions with their typical arguments. Generally, the source evolution function, $\theta_z$, will depend only on $z$; the energy analogue to the luminosity function, $\theta_E$, will have a power law dependence on $E_{\nu_e}$ described by index $\gamma$ (where broken power laws are used, $\gamma$ is subscripted accordingly) up to a hard cutoff at the maximum spectral energy $E_{\nu_e,\text{max}}$; and the spectrum, $\theta_{\nu_e}$, will have a power law dependence on $\nu_e$ described by index $\alpha$ (and $\beta$ where broken power laws are used).

Following the work of \cite{macquart_frb_2018}, the intrinsic event rate energy function can be related to the differential observed rate with fluence via
\begin{equation}\label{eq:drdfSmooth}
    \frac{dR}{dF}=\int dz\; 16\pi^2D_c^2D_L^2\frac{1}{(1+z)^3}\frac{dD_c}{dz}\Theta_E(z, E_{\nu_e}, \nu_e),
\end{equation}
where $D_L$ and $D_c$ are the comoving and luminosity distances respectively --- for a complete derivation see appendix \ref{app:derivations:SmoothDiffRate}.

\subsection{Rates in a Clumpy Universe}\label{subsec:clumpyRates}
In an inhomogeneous universe the relation between observed fluence and emitted energy is more complicated. Naturally it is dependent on the total magnification of the source. However as our magnification is with respect to the empty beam case, the luminosity distance cannot be calculated for a smooth universe as in \S \ref{subsec:smoothRates}. Instead it must be expressed as a function of $D_\eta$ via Etherington's reciprocal relationship ($D_L = D_\eta(1+z)^2$) \citep{etherington_lx_1933}. 

The probability of any given line of sight to a source at redshift $z$ having magnification $\mu$ is given by the PDF described in Eq. \ref{eq:PDF}. For most of our calculations we assume the lensing is well characterised by geometric optics. At radio wavelengths this assumption may break down as we show in \S \ref{app:WaveEffects}.

Combining these elements as elaborated in the derivation in appendix \ref{app:derivations:ClumpyDiffRate} the differential rate with fluence for transients in an inhomogeneous universe with a smooth matter fraction $\eta$ is
\begin{align}
    \frac{dR}{dF}=\int dz&\; 16\pi^2D_c^2\left(D_\eta(1+z)^2\right)^2\frac{1}{(1+z)^3}\frac{dD_c}{dz}\nonumber\\
                        \times&\int d\mu\; p(\mu, z)\Theta_E(z, E_{\nu_e}, \nu_e)\frac{1}{\mu}\label{eq:drdfClumpy}\:.
\end{align}

\subsection{Numerical Implementation}\label{subsec:numImplementation}
To evaluate the differential rates we used SciPy's implementation of the \textsc{fortran} quad pack numerical integration. For the smooth universe calculation of $dR/dF$ we integrate in log space over the domain [$z_{\text{min}}$, $z_{\text{max}}$]. In our physical picture the calculated value $dR/dF$ then corresponds to a hollow sphere between redshifts [$z_{\text{min}}$, $z_{\text{max}}$]. 

In line with the expectations in a LambdaCDM universe, we assume that on large scales the Universe is homogeneous, hence we set a minimum redshift condition of $z_{\text{min}}=0.001$ (D$\sim$ 4 Mpc), corresponding to the scale between galaxies. We do not model the contribution to the event rate from below this scale, as the local structure of our Universe would need to be accounted for. Even with $z_{\text{min}}$ set at 0.001, our hollow sphere still well approximates the volume of filled sphere out to $z_{\text{max}}$. 

The upper redshift boundary corresponds to the designated spatial distribution, e.g. for $\theta_z\propto$ cosmic star formation rate \citep[CSFR; throughout this paper we make use of the CSFR outlined in ][]{madau_cosmic_2014} we set $z_{\text{max}}=100$ where star formation is negligible.

In the case of a clumpy universe our redshift integration must be restricted to a higher minimum, $z_{\text{min, lensed}}$, to ensure stable integration. As seen in Eq. (\ref{eq:meanMagnification}), $\langle\mu\rangle$ is dependent upon redshift and the smooth matter fraction, $\eta$. For high values of $\eta$ and low redshifts $\langle\mu\rangle-1$ will be small. To have a magnification PDF of the form of Eq. (\ref{eq:PDF}) with a small mean magnification requires that the PDF be extremely concentrated around $\langle\mu\rangle$. For mean magnifications $\langle\mu\rangle-1\lesssim 10^{-5}$ this peak can be missed in the integration domain, destroying the validity of the result. For inhomogeneous universes ($\eta<1$) we set $z_{\text{min, lensed}}\geq z_{\text{min}}=0.001$ to ensure a valid result. For lower values of $\eta$, $z_{\text{min, lensed}}$ is decreased such that the minimum $\langle\mu\rangle$ remains constant. Given the extremely low mean magnification in this low redshift regime the clumpy and smooth universe results are unlikely to vary significantly. Therefore, when $z_{\text{min, lensed}}>z_{\text{min}}$ we add the smooth universe result over the domain [$z_{\text{min}}$, $z_{\text{min, lensed}}$] to the clumpy integral result to keep the smooth and clumpy $dR/dF$ results consistent.

For the clumpy universe calculation the inner integral in Eq. \ref{eq:drdfClumpy} is performed in log space over the transformed domain of $\Delta\mu=\mu-1$ to aid numerical integration by spreading out the sharply varying behaviour of the magnification PDF over a greater dynamic range. This integration is performed over the domain [$\Delta\mu_{\text{min}}$, $\Delta\mu_{\text{max}}$], where $\Delta\mu_{\text{min}}=10^{-15}$ as restricted by float precision \footnote{Technically the float precision of the exponential term in Eq. (\ref{eq:PDF}) is violated significantly before $\Delta\mu=10^{-15}$, however the impact on the accuracy of the result is negligible.} and $\mu_{\text{max}}=10^{20}$, beyond which we expect negligible contribution to the integration. 

\section{Fractional Change Due to Lensing}\label{sec:fractionalChange}

Here and in following sections we illustrate $dR/dF$ using plausible values of the population functions for FRBs \citep{james_fast_2021, luo_frb_2020}, and vary individual parameters over a broad range of typical values.

\begin{figure}
    \centering
    \includegraphics[width=0.5\textwidth]{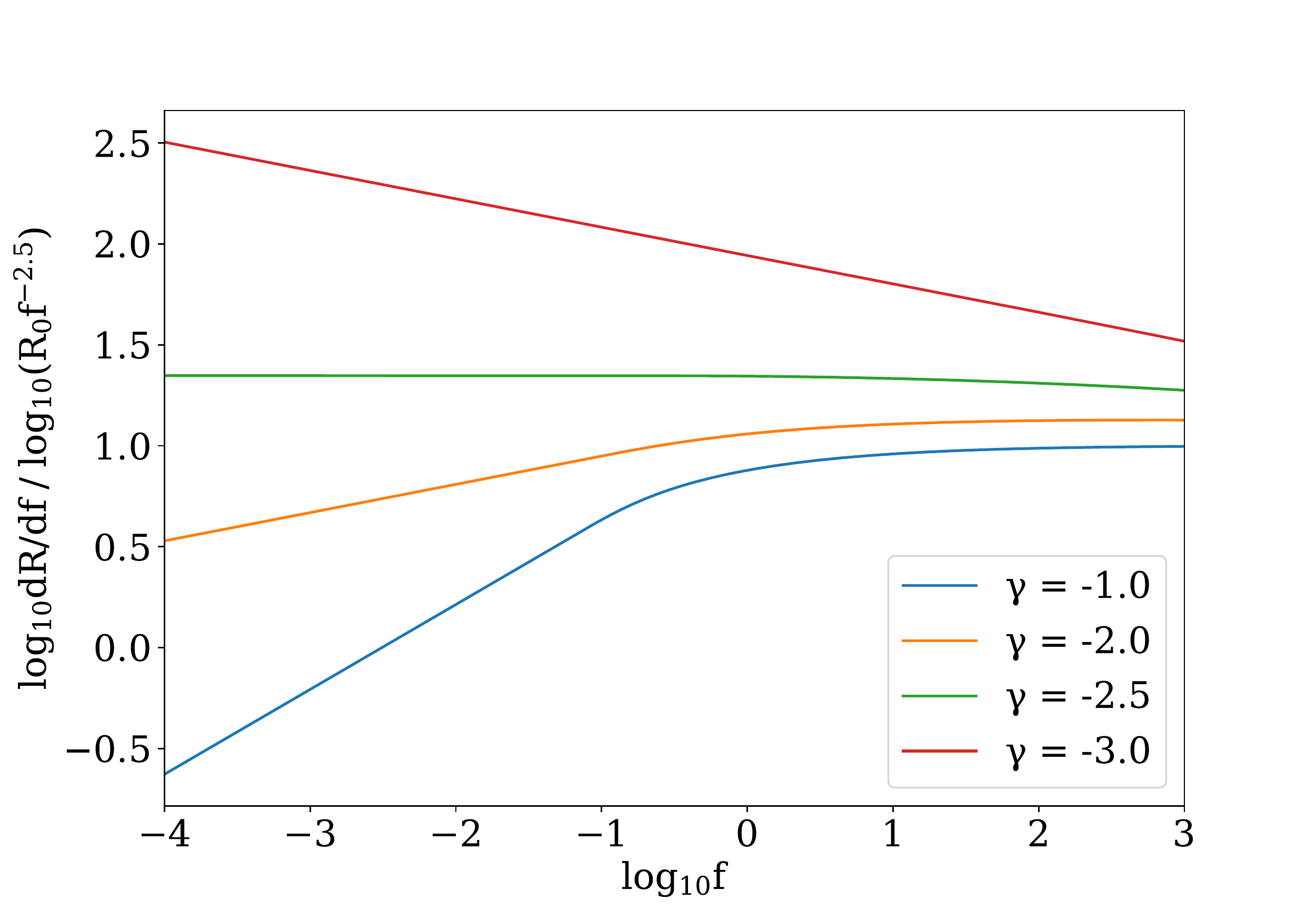}
    \caption{Observed differential event rate as a function of normalised fluence in a smooth universe for an event rate energy function with $\alpha=-1.0$, $E_{\nu_e, \text{max}}=10^{33}$erg/Hz, a uniform comoving spatial density and a range of $\gamma$ values given in the legend and $z_{\text{max}}=100.0$. We normalise the result to the Euclidean expectation given by $R_0f^{-2.5}$.}
    \label{fig:Generic}
\end{figure}

Fig. \ref{fig:Generic} plots $dR/df$ as a function of $f$ for a smooth universe, where $f$ is the observed fluence normalised to what would be observed for an $E_{\nu_e, \text{max}}$ burst at redshift $z=1$ in a smooth universe. We normalise the rates to the expected $dR/df$ for a Euclidean universe, $R_0f^{-2.5}$ where $R_0$ is the rate density in the local universe.

Fig. \ref{fig:Generic} shows that $dR/df$ in a smooth universe has roughly two fluence domains of behaviour for event rate energy functions with $\gamma>-2.5$. We define these regions about a break fluence $f_{b,1}=10^{-1.16}$ which is the fluence corresponding to $E_{\nu_e,\text{max}}$ at $z_{\text{max}}$. At the high fluence end ($f>f_{b,1}=10^{-1.16}$) the maximum redshift is determined by $E_{\nu_e,\text{max}}$. Because the power law index of the energy function, $\gamma>-2.5$ is shallower than the expected Euclidean evolution $\propto f^{-2.5}$ the change in rate with fluence is dominated by the change in sources due to a restricted redshift, as opposed to having fewer bursts at higher energies. Therefore $dR/df$ in a smooth universe has a power law index at the high fluence end approaching the Euclidean expectation. 

On the low fluence end ($f<f_{b,1}=10^{-1.16}$), where $z_{\text{max}}$ is instead restricted by the spatial distribution $\theta_z$, a change in the observed fluence does not affect $z_{\text{max}}$ and the change in the observed rate is dominated by seeing fewer bursts at higher energies. Therefore at the low fluence end $dR/df$ adopts power law index seen in the energy function of $\gamma$. If the energy function has a steeper index, $\gamma\leq-2.5$ then for a uniform spatial distribution the change in the number of bursts due to the energetics will dominate across all fluences and the Euclidean behaviour will never be recovered. 

In a clumpy universe $dR/df$ depends on the convolution of the intrinsic energy function with $p(\mu)/\mu$ as per Eq. \ref{eq:drdfClumpy}. For gravitational lensing this convolution kernel is generally $\propto\mu^{-4}$ in the high magnification limit. Therefore, for all energy functions with $\gamma>-4$\footnote{We only calculate $dR/df$ in a clumpy universe for $\gamma>-4$. Because $p(\mu)/\mu\propto\mu^{-4}$ at high $\mu$, for $\gamma\leq -4$, the inner integral of Eq. \ref{eq:drdfClumpy} would not converge.} this convolution will be dominated by the intrinsic energy function, and the behaviour of $dR/df$ will be well approximated by the smooth universe behaviour described above. The exception will be the case where all events are highly magnified as we shall discuss in \S \ref{sec:AllLensed} and the edge effects we describe below.

\begin{figure}
    \centering
    \includegraphics[width=0.5\textwidth]{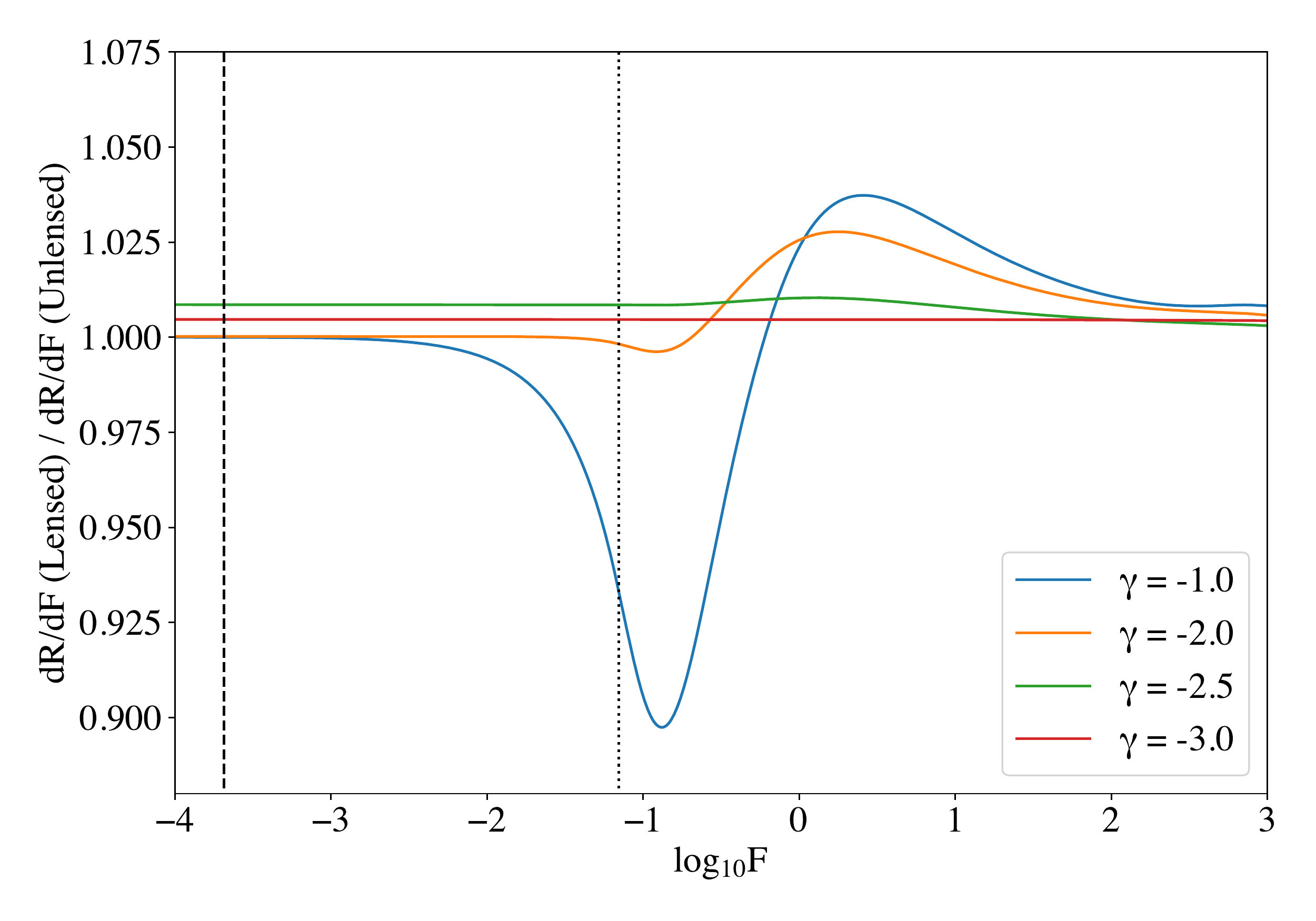}
    \caption{Differential rates for a clumpy universe from Fig. \ref{fig:Generic} normalised by the corresponding values in a smooth universe. Dashed and dotted lines denote the break fluences $f_{b,\eta}$ and $f_{b,1}$ for clumpy and smooth universes respectively.}
    \label{fig:GenericFractional}
\end{figure}

To discern the effect of lensing we express our results as the differential event rate ($dR/df$) in a clumpy universe with a smooth matter fraction $\eta<1$, normalised by the differential event rate in a smooth universe ($\eta=1$). Fig. \ref{fig:GenericFractional} shows the $\eta=0$ case, depicting a $1-10\%$ fractional change in $dR/df$ due to lensing that has a characteristic shape common to all values of $\gamma$. 

Fig. \ref{fig:GenericFractional} shows the low fluence regime for all values of $\gamma$ and all fluences for $\gamma\leq-2.5$ to have approximately constant fractional difference between the lensed and unlensed differential rates. $dR/df$ in these regions are dominated by the energetics and hence show similar behaviour despite any lensing, consistent with our expectation.
 
The fluctuation structure is comprised of an initial decrease, before a sharp increase which then tends back towards unity. To understand why this structure appears we must look to the break fluence $f_b$. For $f>f_b$, the maximum redshift becomes restricted by $E_{\text{max}}$ and $dR/df$ becomes dominated by a reduction of the volume out to which sources can be observed as discussed previously. In a smooth universe this occurs at $f_{b,1}=10^{-1.16}$, shown as a dotted line in Fig. \ref{fig:GenericFractional}. In a clumpy universe however this occurs at a lower fluence of $f_{b,\eta}=10^{-3.69}$, shown as a dashed line in Fig. \ref{fig:GenericFractional}. The lower fluence is due to a demagnification of $1/\langle\mu\rangle$ (from Eq. (\ref{eq:meanMagnification}),  $D_1^2(z_\text{max}=100)/D_0^2(z_{\text{max}}=100)=0.00295\approx10^{-3.39}/10^{-1.16}$) associated with viewing along an empty beam in a clumpy universe compared to a smooth one. Therefore, as we increase fluence $dR/df$ becomes dominated by the reduction to $z_{\text{max}}$ in a clumpy universe before it does so for a smooth universe, resulting in the dip. Once we hit $f=10^{-1.16}$, the smooth universe also becomes dominated by reduction to $z_\text{max}$ resulting in an inflection point in the fractional change in accordance with $z_{\text{max}}$ decreasing faster in a smooth universe. As the fluence increases and the maximum redshift approaches the nearby universe, the mean magnification decreases and clumpy and smooth universes become indistinguishable, resulting in $dR/df$ values that converge as seen on the high fluence end of Fig. \ref{fig:GenericFractional}.

\section{Intrinsic Parameter Variation}\label{sec:IntrinsicParamVar}
Apart from $\gamma$, our model takes in a number of input parameters as discussed in \S \ref{subsec:smoothRates}. Below we characterise the response of $dR/df$ to variation of these parameters.

\subsection{Spatial Distributions}\label{subsec:spatialDist}

\begin{figure}
     \centering
     \includegraphics[width=0.5\textwidth]{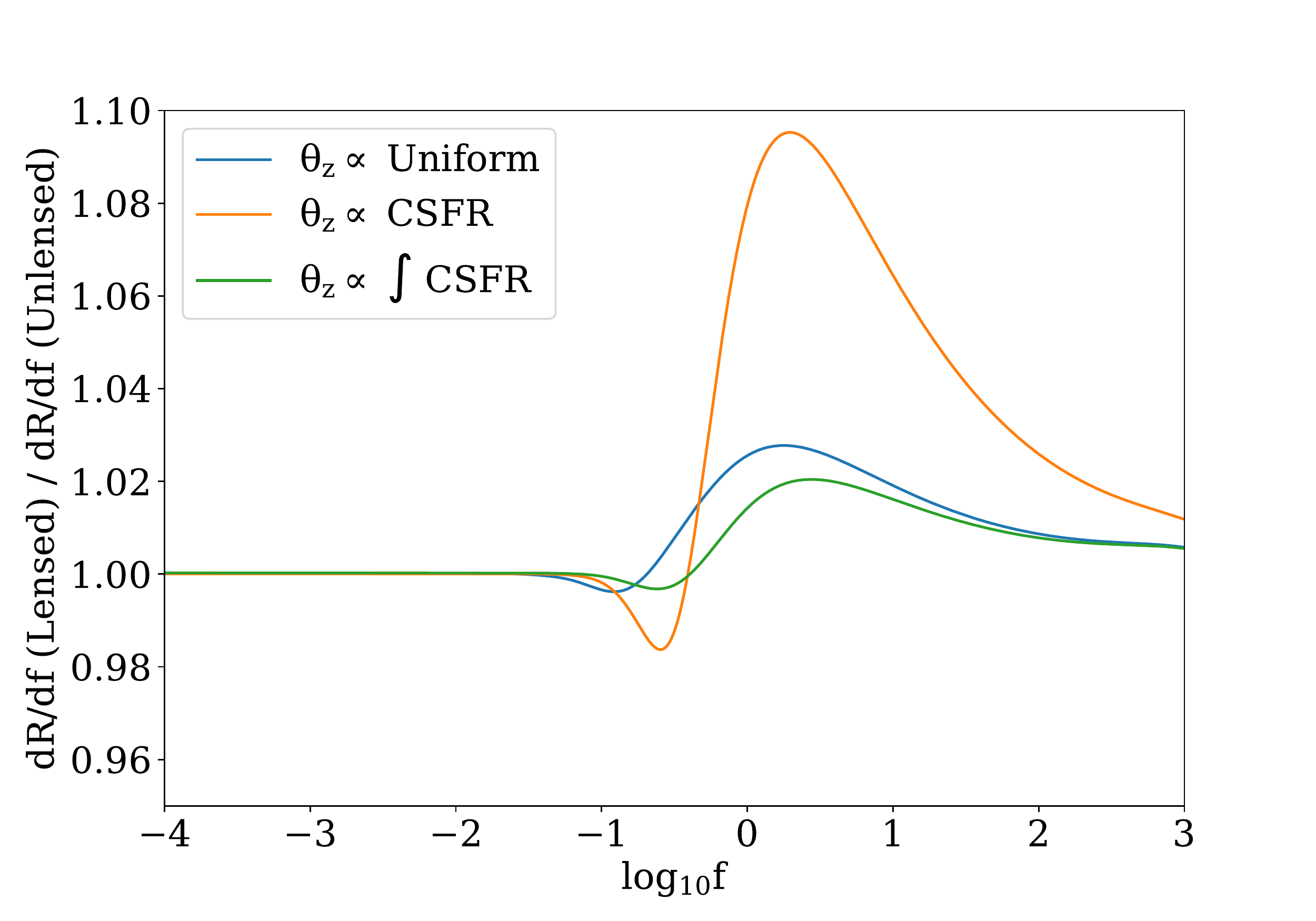}
     \caption{Differential rates for a clumpy universe normalised by their smooth universe equivalents for a selection of spatial distributions $\theta_z$. Other parameters of the population functions are identical to those in Fig. \ref{fig:Generic} with $\gamma=-2.0$}
     \label{fig:GenericSpatial}
 \end{figure}

 Our calculations so far have been restricted to an intrinsic rate with a uniform comoving spatial distribution. More realistically $\theta_z$ is likely to be proportional to some integral over the CSFR, owing to the stellar origin of the extreme environments that produce (or are candidate progenitors for) many extragalactic transients \citep{gehrels_gamma-ray_2009, platts_living_2019}. For progenitors that are short lived this integral will be over a small portion of the CSFR and hence the result proportional to the CSFR itself. Conversely, progenitors that emit bursts over a time period comparable to the age of the universe will have $\theta_z$ proportional to the current number of stars, i.e. the CSFR integrated over all redshifts above $z$. Taking $\gamma=-2.0$ we plot $dR/df$ for these spatial distributions in Fig. \ref{fig:GenericSpatial}. 
 
 Fig. \ref{fig:GenericSpatial} shows that for spatial distributions proportional to the CSFR or its integral, the minimum fractional change occurs at a higher fluence than for a uniform spatial distribution. This is because the CSFR shows a gradual decline beyond $z\sim2$ rather than a hard edge at $z\sim100$. Also of note is that the fractional increase in $dR/df$ from lensing is much larger for the case of $\theta_z\propto$ CSFR than other spatial distributions.
 
\begin{figure}
     \centering
     \includegraphics[width=0.5\textwidth]{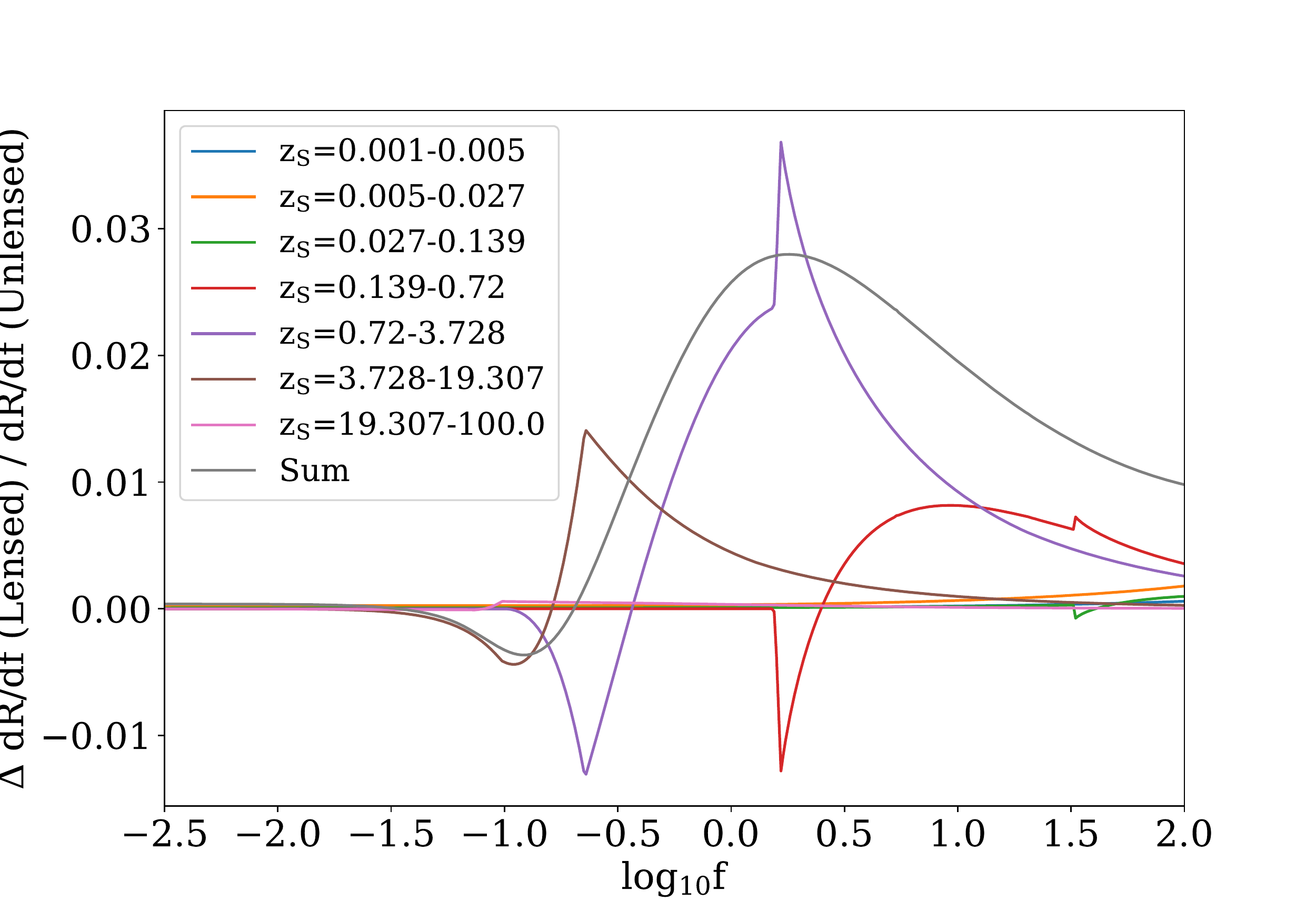}
     \caption{Components in redshift space of fractional change in $dR/df$ due to lensing in a $\eta=0.0$ universe. Components depicted in logarithmically spaced redshift bins. Sum is equivalent to those shown in Fig. \ref{fig:GenericFractional} and \ref{fig:GenericSpatial} for a uniform spatial distribution with $\gamma=-2.0$. The sharp peaks seen in the curves representing each redshift bin are the result of constructing hard bin boundaries in redshift space.}
     \label{fig:GenericRedshiftDecomp}
 \end{figure}

 To see why, it is informative to decompose the fractional change due to lensing for a uniform spatial distribution into its components in redshift space. Fig. \ref{fig:GenericRedshiftDecomp} shows that in the case of a $\theta_z$ which is uniform in comoving space, a majority of the fractional increase due to lensing comes from the $z=0.72-3.728$ region. This is the same region in redshift space where the CSFR peaks and hence for $\theta_z\propto$ CSFR the rate of bursts coincidentally peaks where the fractional change due to lensing is greatest, enhancing the effect of lensing.   
 
 \subsection{Spectral Indices}\label{subsec:spectralIndices}
\begin{figure}
     \centering
     \includegraphics[width=0.5\textwidth]{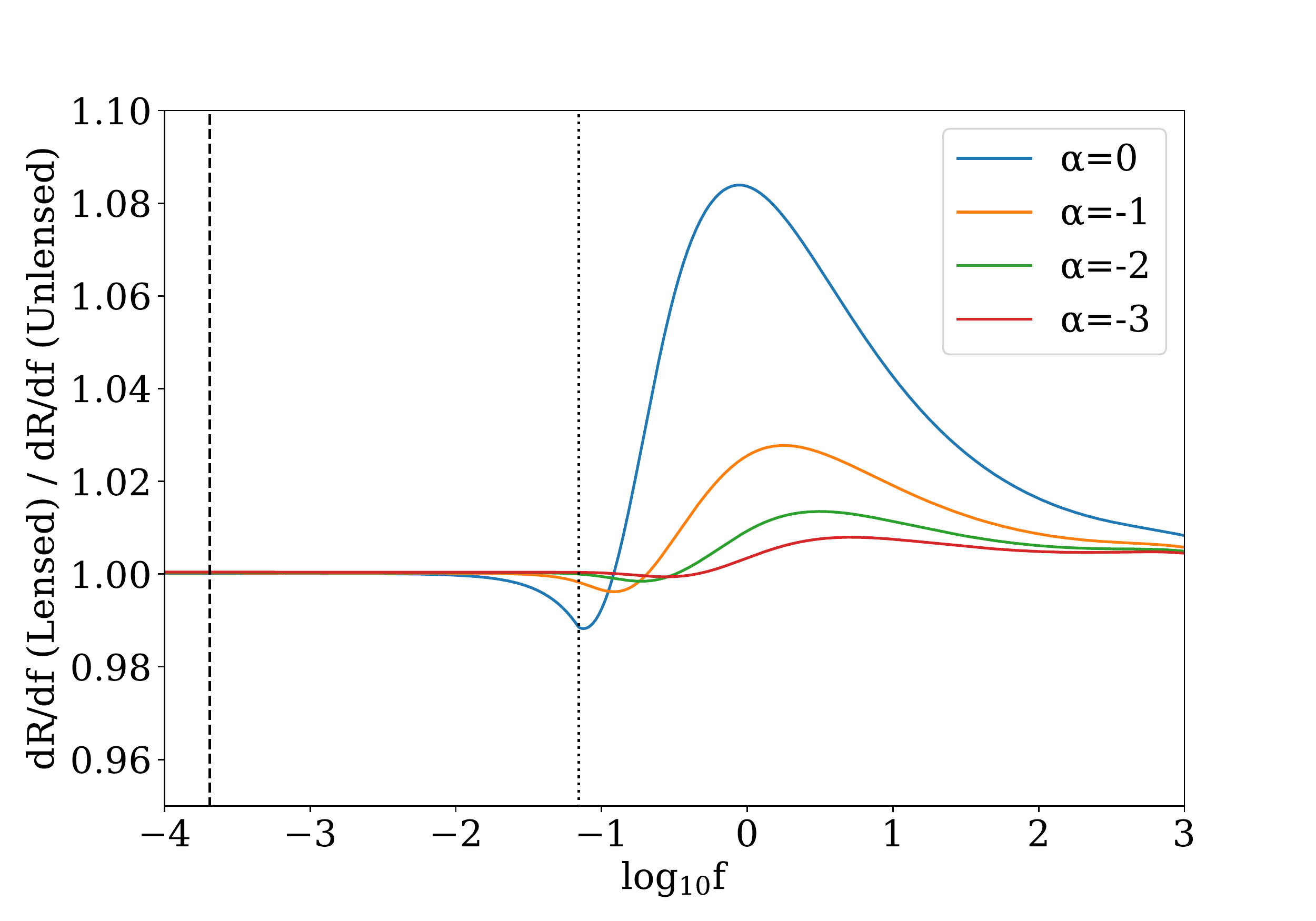}
     \caption{Differential rate in an $\eta=0$ universe normalised by the smooth universe equivalent. We select our parameters to be the same as Fig. \ref{fig:Generic} with $\gamma=-2.0$ but for a variety of $\alpha$ values.}
     \label{fig:GenericFractionalSpectral}
 \end{figure}

  Fig. \ref{fig:GenericFractionalSpectral} shows that the fractional change due to lensing varies with the spectral index $\alpha$ similarly to the energy index $\gamma$. Steep, negative spectral indices rapidly decrease the burst rate at high frequencies which suppresses the burst rate at higher redshifts where the emission frequency associated with any given observed frequency is higher by a factor of $(1+z)$. Because all significant lensing effects occur in the distant universe where the mean magnification is higher, a suppression to the intrinsic rate at high redshifts restricts the effect of lensing as shown.

\subsection{$E_{\text{max}}$}

\begin{figure}
    \centering
    \includegraphics[width=0.5\textwidth]{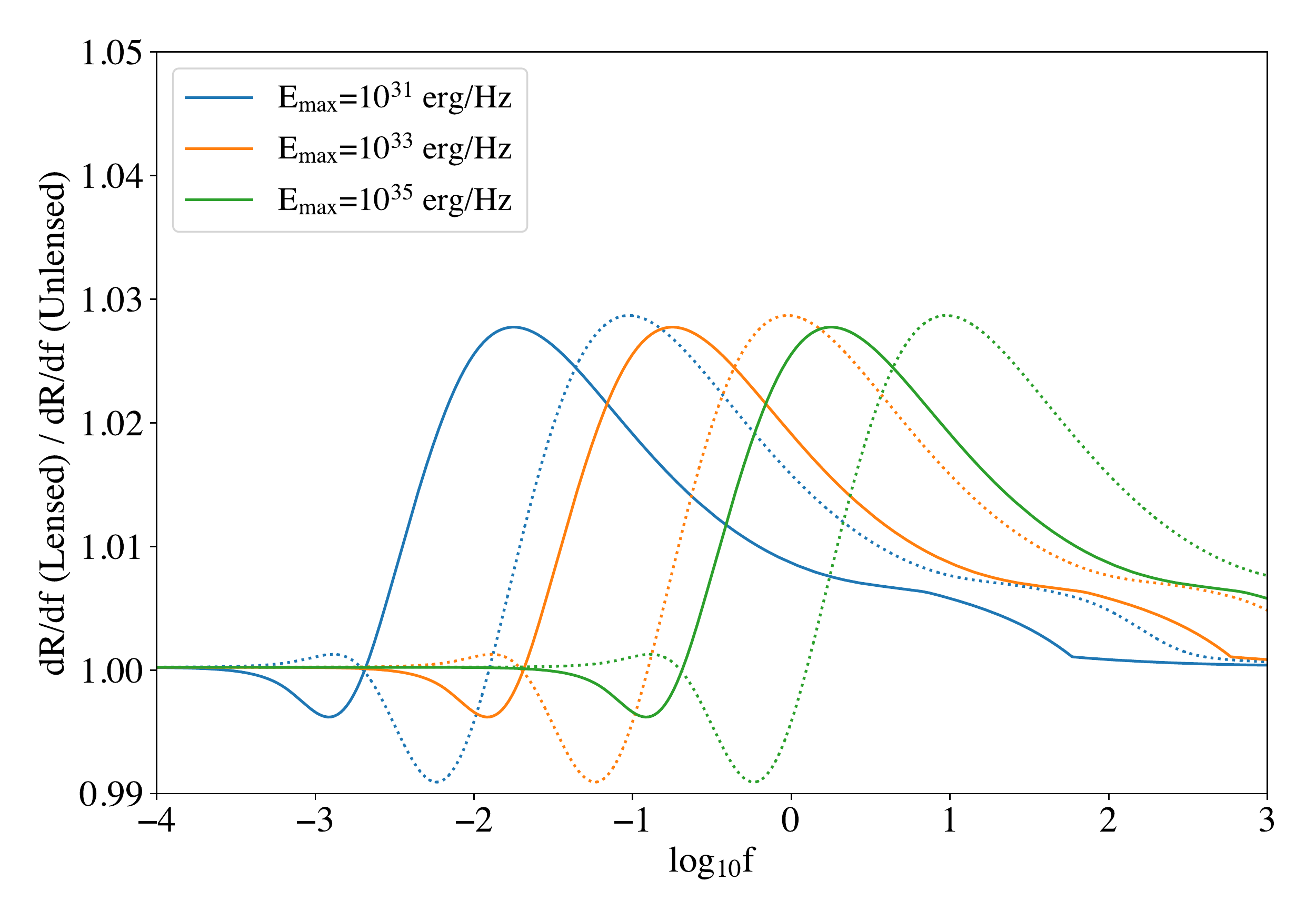}
    \caption{As per Fig. \ref{fig:Generic} with $\gamma=-2.0$ but a varying $E_{\nu,\text{max}}$. Dotted lines represent the case of an exponential cutoff in $\theta_E$ above $E_{\nu,\text{max}}$ as opposed to a hard boundary.}
    \label{fig:GenericFractionalEmax}
\end{figure}

The maximum energy of a burst $E_{\nu_e,\text{max}}$ defines where $f_b$ will lie and so will affect where the structure in the above figures will be in fluence. As shown in Fig. \ref{fig:GenericFractionalEmax} increasing or decreasing $E_{\nu_e,\text{max}}$ shifts the fluctuations in fractional change linearly with $f$. Apart from translation, the impact of changing $E_{\nu_e,\text{max}}$ is negligible.

The shape of the structures seen in Fig. \ref{fig:GenericFractionalEmax} is caused somewhat by the sharpness of the hard cutoff at $E_{\nu_e, \text{max}}$. Because a hard cutoff is a rather unrealistic feature of an energy function we have also calculated Fig. \ref{fig:GenericFractionalEmax} for an exponential cutoff at the same boundary. The results of this calculation are also contained in Fig. \ref{fig:GenericFractionalEmax} as the dotted lines. The figure shows only mild differences from a hard cutoff, including a short rise before deeper decreases, each structure is also shifted to higher fluences. The small scale of these changes shows that the hard cutoff in energy we use is a good approximation for a more realistic sharp decrease in rate beyond $E_{\nu_e,\text{max}}$.

 \subsection{Cosmology}\label{subsec:cosmology}
\begin{figure}
    \centering
    \includegraphics[width=0.5\textwidth]{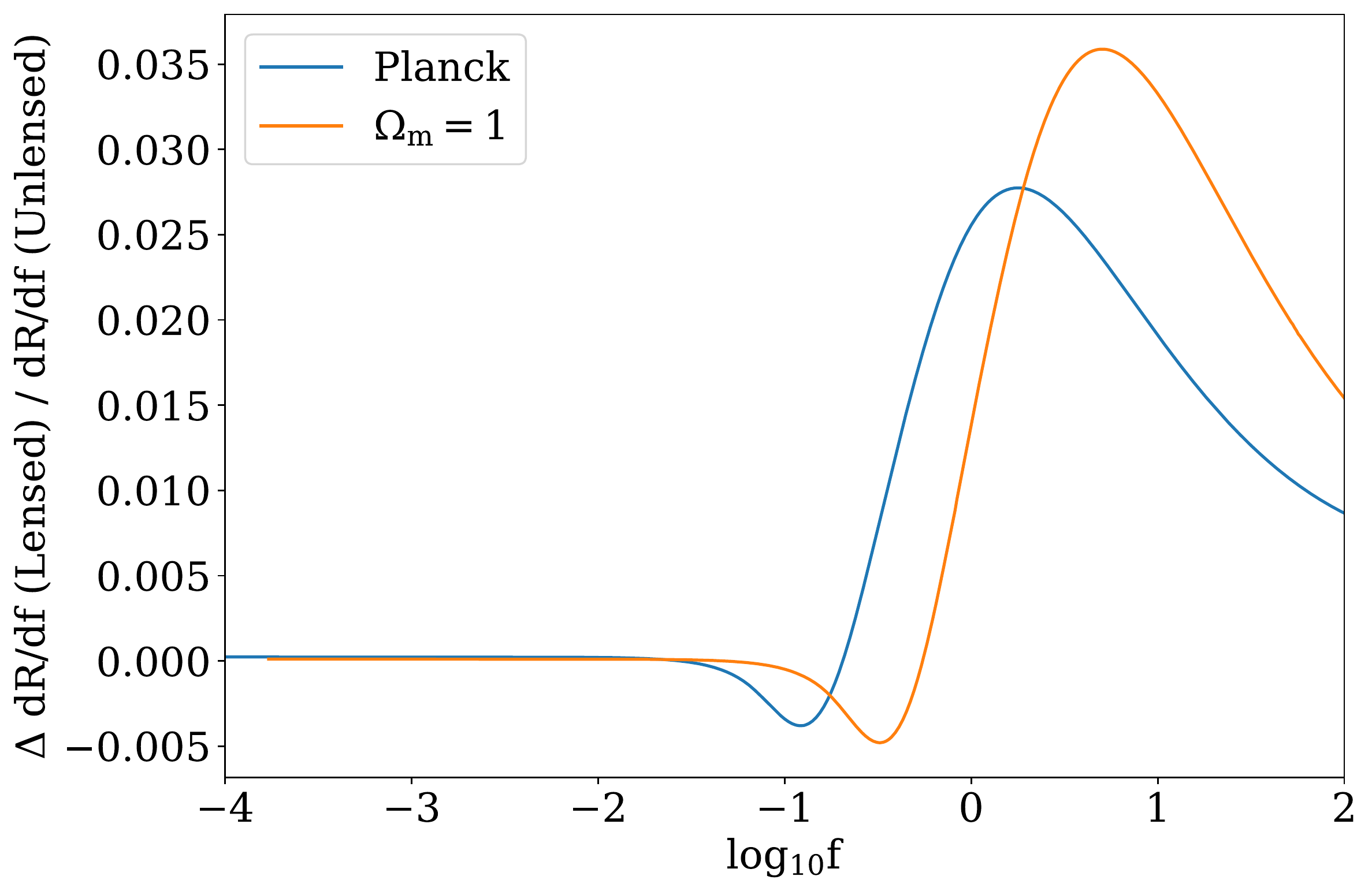}
    \caption{Comparison of the fractional change in $dR/df$ due to lensing between a Planck cosmology and an Einstein De-Sitter universe (i.e. a universe with $\Omega_m=1$). Other parameter chocies are as seen in Fig. \ref{fig:Generic} with $\gamma=-2.0$}
    \label{fig:GenericCosmology}
\end{figure}

    Our results will also depend upon the choice of cosmology used in the model. To demonstrate how changes in cosmology will affect the results we compare the Planck cosmology to the extreme case of an Einstein De-Sitter cosmology that has zero cosmological constant and all of its energy density contained in matter ($\Omega_m=1$). The Hubble constant for each is that given by the Planck constraints. Both are calculated for the case where $\eta=0$, giving a density in lenses of $\Omega_L$=$\Omega_m$. Fig. \ref{fig:GenericCosmology} shows the fractional change due to lensing for each of these choices. It shows only a mild difference between the two cosmologies, with the Einstein De-Sitter universe having a greater peak. Fig. \ref{fig:GenericCosmology} also shows that the fractional fluctuation in $dR/df$ due to lensing in an Einstein De-Sitter universe is shifted to a higher fluence. This is expected as an $\Omega=1$ universe has a lower luminosity distance at a given redshift, giving a commensurately higher $f_b$ in both the lensed and unlensed case.    

\section{Are All FRBs Lensed ?}\label{sec:AllLensed}

Above we have describe the scenario where a burst population which could be observed in a smooth universe is altered by lensing. However, an alternative scenario which has been discussed in the FRB community is the possibility that bursts are only observed because of gravitational lensing, i.e. all observed bursts are intrinsically low energy but highly magnified. In such a situation the minimum redshift becomes important. We define $f_{\text{max}}$ as the largest fluence where an observer will see a burst with $\mu=1$, corresponding to $E_{\nu_e,\text{max}}$ at $z_{\text{min}}$. For $f>f_\text{max}$ only magnified bursts are observed, and because the entire spatial domain contributes, the behaviour of $dR/df$ with fluence will be determined entirely by the inner integral over magnification in Eq. (\ref{eq:drdfClumpy}). As $f$ increases, the intrinsic energy required at each magnification will increase and the observed rate will decrease following $\theta_E$ with index $\gamma$ (fewer bursts at higher energies), assuming again that $\gamma>-4$. Additionally, the minimum observed magnification will increase with $f$, resulting in a decrease to $dR/df$. If the PDF with a factor $1/\mu$ behaves as $p(\mu)/\mu \propto \mu^{\xi-1}$ (as it does in the high magnification limit) the integration from $[\mu_{\text{min}},\infty]$ will vary with $\mu_\text{min}$ following a power law of index $\xi$. Therefore as $f$ increases and $\mu_{\text{min}}$ increases, $dR/df$ will also vary with index $\xi$. These behaviours will occur simultaneously, however if one is significantly steeper we expect that it will dominate the change to $dR/df$, e.g. for $\xi \ll \gamma$, $dR/df$ will be approximately $\propto f^\xi$, and vice versa. 

\begin{figure}
    \centering
    \includegraphics[width=0.5\textwidth]{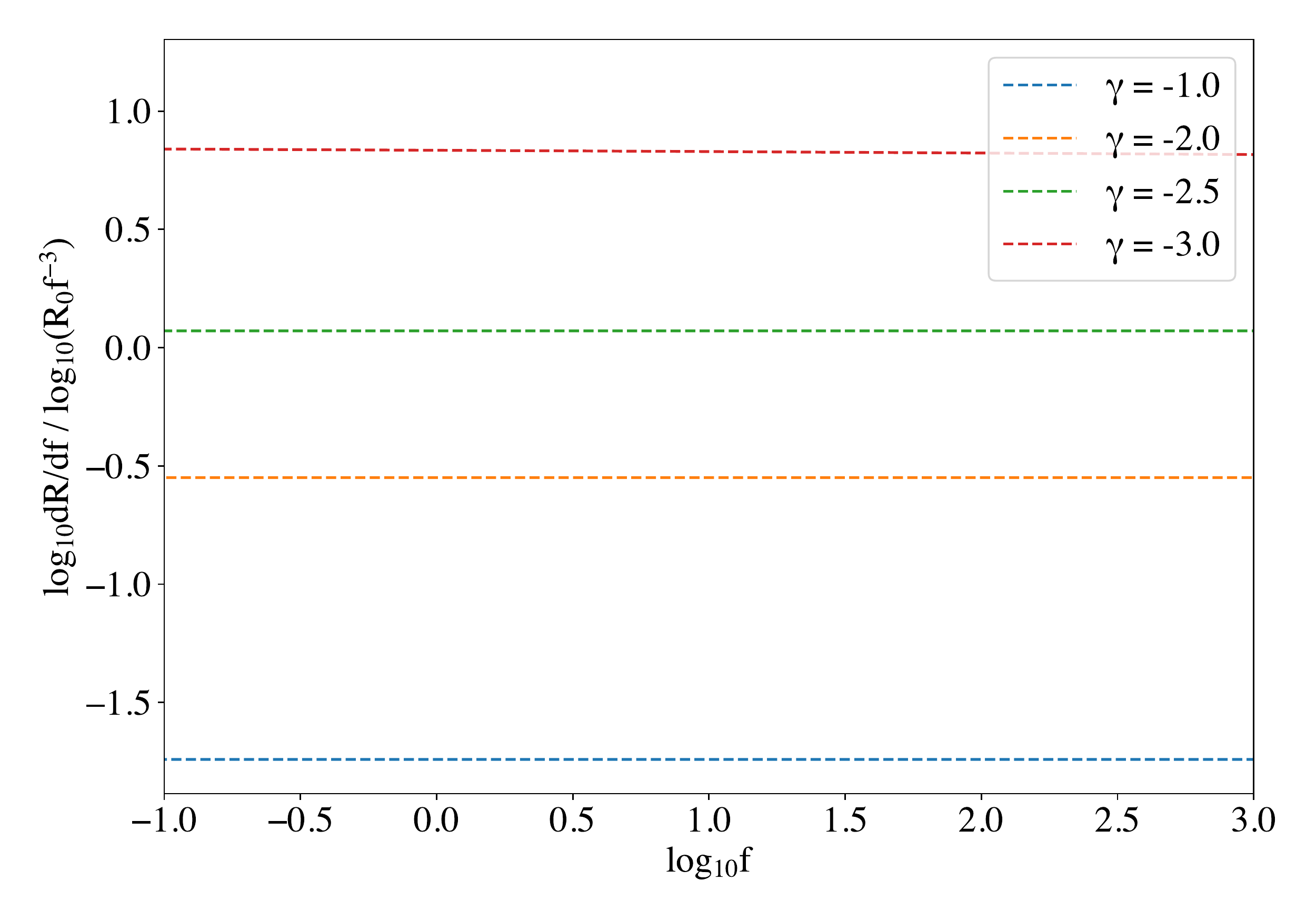}
    \caption{Differential rate as a function of normalised fluence in a clumpy universe with $\eta=0$, for an event rate energy function with $\alpha=-1$, $E_{\nu_e,\text{max}}=10^{26}$ erg/Hz, a uniform comoving spatial density and a range of $\gamma$ values given in the legend. As specified in \S \ref{subsec:numImplementation}, $z_{\text{min}}=0.001$ (D$\sim$ 4 Mpc), corresponding to the scale between galaxies. We normalise the result to the same Euclidean expectation used in Fig. \ref{fig:Generic} and use the same normalisation energy in $\theta_E$. $f_{\text{max}}\approx 10^{-6}$}
    \label{fig:GenericLensedAlt}
\end{figure}

Given that in the geometric optics limit for a stationary universe our chosen PDF varies as $p(\mu)\propto\mu^{-3}$ the expected behaviour for all $\gamma\geq-3$ is to have $dR/df$ vary with a power law index of $-3$. Fig. \ref{fig:GenericLensedAlt} shows precisely this scenario, plotting $dR/df$ for the same selection of $\theta_E$ functions used earlier but with $E_{\nu_e,\text{max}}=10^{26}$ erg/Hz which is more in line with the spectral energy observed for the Galactic FRB \citep{the_chimefrb_collaboration_bright_2020}. Fig. \ref{fig:GenericLensedAlt} shows that for all $\gamma\geq-3$ the $dR/df$ has an index of $-3$ as expected. 

If all FRBs were highly magnified in a stationary universe, the behaviour of $dR/df$ would be consistent with a $\gamma=-3$ intrinsic energy function in a smooth universe. Estimates of $\gamma$ outside the context of lensing would therefore yield $\gamma=-3$. Best estimates of $\gamma$ from observed FRBs give $\gamma\approx-2$ \citep{james_fast_2021, luo_frb_2020}, which never has $dR/df$ behaviour consistent with a $\gamma=-3$ model and therefore allows us to refute a scenario where FRBs are only observable due to high magnifications from stationary gravitational lenses.

\begin{figure}
    \centering
    \includegraphics[width=0.5\textwidth]{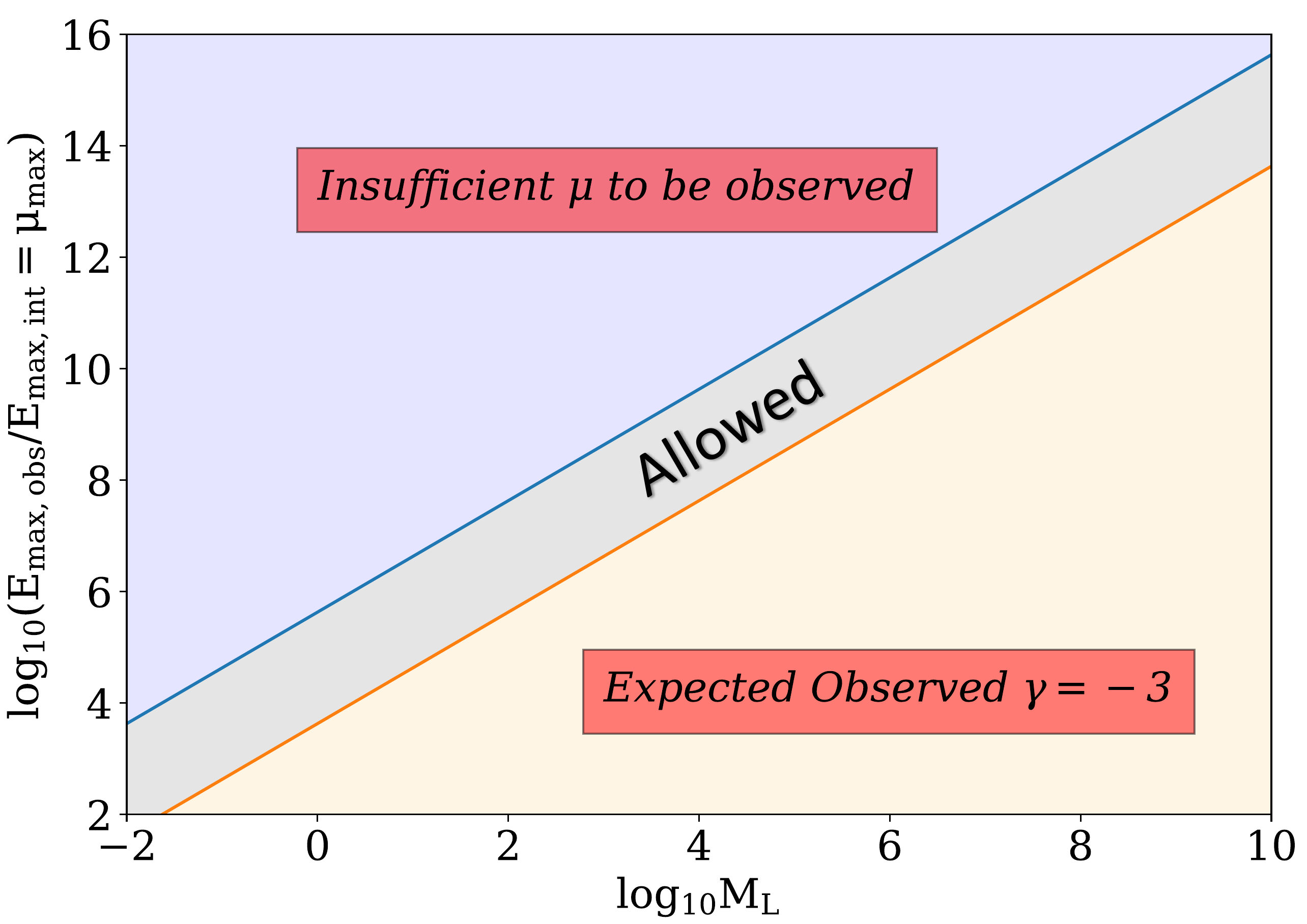}
    \caption{Range of lens masses excluded from highly magnifying all FRBs. Blue line is set by the maximum magnification under wave optics for a 1GHz FRB lensed by an $M_L$ mass point lens. Orange line corresponds to the magnification 1.5 orders of magnitude below the maximum. Orange and blue regions highlight parts of the parameter space excluded by observed $\gamma$ or required magnification respectively. Gray region highlights the unconstrained area of the parameter space associated with prominent intereference fringers in the cross section.}
    \label{fig:AltScenCond}
\end{figure}

In appendix \ref{app:WaveEffects} we show that as a result of wave optics low mass lenses may not follow $p(\mu)\propto \mu^{-3}$. Given this potential departure from the $p(\mu)\propto\mu^{-3}$ behaviour we can only use $\gamma\neq-3$ to refute that all FRBs are highly magnified by lenses of certain mass. The range of masses which are constrained depends on the magnification required to make the bursts observable, i.e. the maximum apparent energy normalised by the maximum intrinsic energy $\mu_{\text{max}}=E_{\text{max, obs}}/E_{\text{max, int}}$. For low lens masses the maximum magnification will be insufficient to make low intrinsic energy bursts observable, allowing us to rule them out by default. For higher intrinsic energies and higher lens masses the lensing behaviour will approach the geometric expectation, $p(\mu)\propto\mu^{-3}$, which are then ruled out as FRB $\gamma\neq-3$. We plot these conditions in Fig. \ref{fig:AltScenCond}, assuming that magnifications $\mu<\mu_{\text{max}}/10^{1.5}$ have $p(\mu)\propto\mu^{-3}$ as describe in appendix \ref{app:WaveEffects}. The figure shows the excluded regions for FRBs, with the intermediate region in grey. Here lensing is of sufficient magnification to make bursts observable but close enough to the maximum magnification to have prominent fringes in the cross section that change the behaviour from $p(\mu)\propto\mu^{-3}$. In this region of the parameter space, we cannot rule out that all FRBs are highly magnified on the basis of $\gamma$ alone. 

\section{How Does Lensing Affect Fast Transient Rates}\label{sec:specificCases}
Assuming that both FRB and GRB populations are intrinsically transient, and not all highly magnified, we have shown in appendix \ref{sec:fractionalChange} that any effect on their differential rates from lensing will be small. Therefore, estimates of their intrinsic parameters, i.e. $\gamma$, $\alpha$ and $\theta_z$, made without accounting for the possibility of lensing will approximate the true population parameters well even if all of the Universe's matter were to be contained in lenses \footnote{With the exception of $E_{\text{max}}$ which may be drastically affected by lensing but has little impact on the inferred value other parameters}. It is therefore appropriate to use the observed population parameters as inputs to our model when calculating the expected $dR/df$ for transient populations in a clumpy universe. In this section we calculate $dR/df$ specific to each transient class for universes with varying $\eta$. We display these rates normalised to what a uniform spatial distribution at the local rate would yield, as well as the fractional differences due to lensing.

We model the effect of lensing on FRBs, long GRBs and short GRBs. We use literature values to build fiducial event rate energy functions in each case as discussed in the following sections. We stress that these models are simplified for the purpose of demonstrating the effect of lensing.

\subsection{Short GRBs}\label{subsec:effectsSGRBs}
To model SGRBs we use the empirical redshift distribution of \citet[][see eq. (21)]{sun_extragalactic_2015} as our $\theta_z$. For consistency we'll also make use of the best fit luminosity and spectral functions from \cite{sun_extragalactic_2015}, i.e. a single power law with $\gamma=-1.6$, and a Band energy function \citep{band_batse_1993} with $\alpha=-0.5$ and $\beta=-2.3$ for $\theta_{\nu}$. \cite{sun_extragalactic_2015} takes these luminosities to be isotropic and bolometric, using a $1-10^4$ keV bandwidth. We also impose a hard maximum luminosity at $L_{\text{max}}=10^{51}$ erg/s which corresponds to the upper bound of the typical energy range for SGRBs \citep{davanzo_short_2015}. To convert these luminosity conditions to spectral energies consistent with our model we divide these luminosities by the assumed $\simeq 10^4$ keV bandwidth and assume that as the peak spectral luminosity at a frequency corresponding to a photon energy of $200$ keV. Furthermore we assume all GRBs to have a duration given by the mode of the \textit{Swift} burst width distribution \citep[SGRBs = 0.1s, LGRBs = 20s; ][]{gehrels_gamma-ray_2009}.

\begin{figure}
    \centering
    \includegraphics[width=0.5\textwidth]{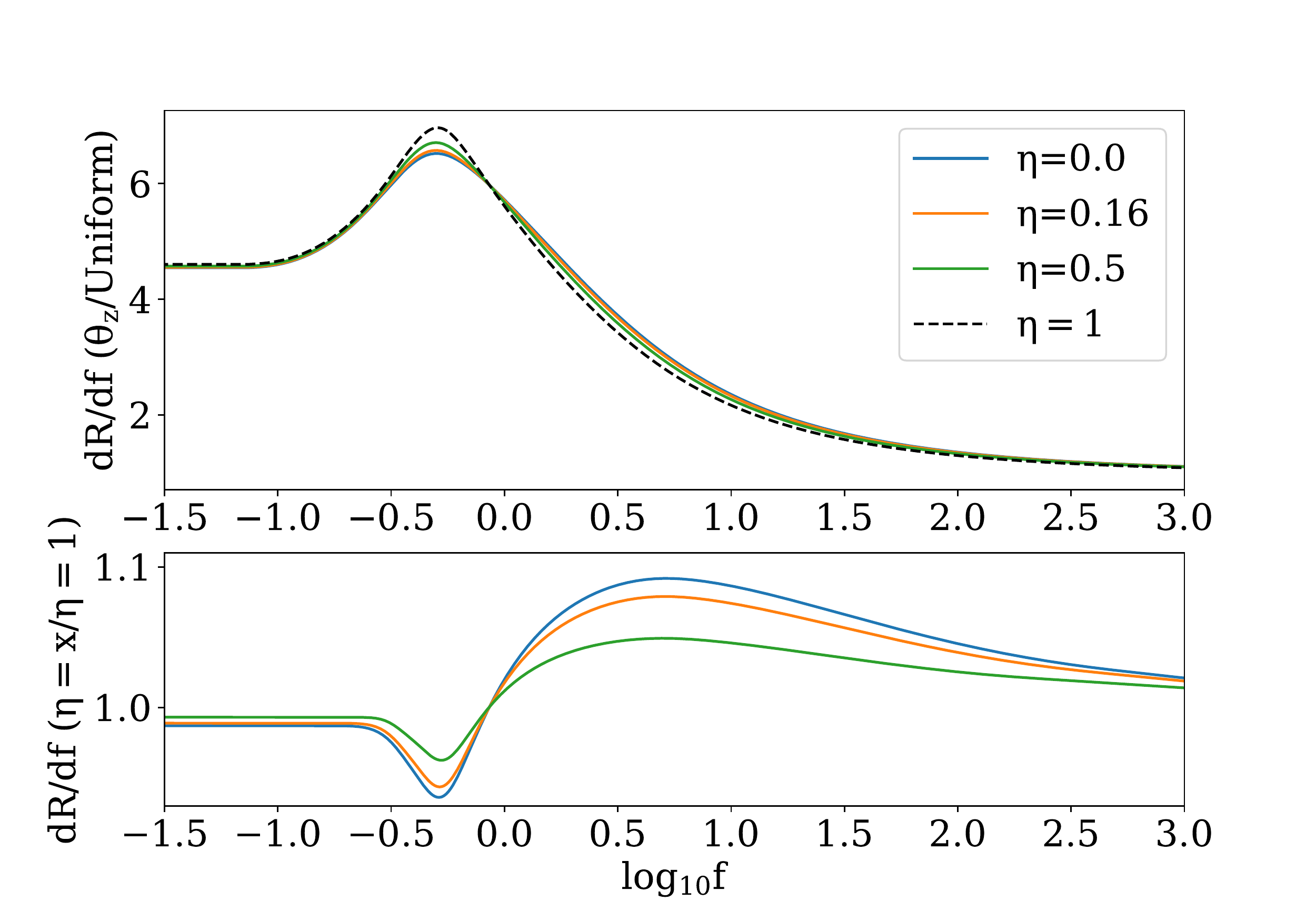}
    \caption{Top panel: $dR/df$ normalised to the differential rate expected for a uniform spatial distribution. Bottom Panel: Fractional change in $dR/df$ due to lensing, i.e. normalised by $dR/df$ in an $\eta=1$ universe. Both panels show results for $\eta$ values of 0, 0.16 and 0.5 the middle of which corresponds to $\Omega_L = \Omega_{\text{DM}}$ in a Planck cosmology.}
    \label{fig:SGRB}
\end{figure}

Fig. \ref{fig:SGRB} shows $dR/df$ calculated for the above SGRB event rate energy function. The top panel normalises the result to $dR/df$ for the same $\theta_E$ and $\theta_\nu$ but a uniform $\theta_z$, whereas the bottom panel shows the result normalised to $dR/df$ for the same $\Theta_E$ in a smooth universe $\eta=1.0$. In both panels we see the high fluence end tending towards 1.0, in line with our expectation for the local rate, which should be the same regardless of the choice of universe or spatial evolution. Moreover the results shown in the top panel can be easily scaled to any choice of normalisation corresponding to different estimates of the local rate of SGRBs (or indeed any of the other transients which we will display similarly). The top panel of Fig. \ref{fig:SGRB} shows the differential rate falling at higher fluences, in accordance with the CSFR decreasing towards lower redshifts. 

Fig. \ref{fig:SGRB} shows that the observed $dR/df$ for our representative SGRB model can fluctuate up to $\approx 10\%$ due to lensing with the scale of these fluctuations decreasing linearly with increasing $\eta$.

\subsection{Long GRBs}\label{subsec:effectsLGRBs}
Those GRBs with a duration above $\sim 2$s are categorised as long (LGRB) and originate from core collapse supernovae explosions (ccSNe). Given the short lifetime of stars which produce ccSNe, LGRBs should trace star formation closely and hence we model $\theta_z\propto$ CSFR. Observations of LGRBs constrain their luminosity function to be a triple power law with indices $\gamma_1=-1.7$, $\gamma_2=-1.0$ and $\gamma_3=-2.0$ in the respective zones between two break luminosities $L_{b,1}=10^{51}$ erg/s and $L_{b,2}=7.8\times10^{52}$ erg/s \citep{sun_extragalactic_2015}. We also impose a hard maximum luminosity at $L_{\text{max}}=10^{54}$ erg/s which corresponds to the highest luminosity LGRB observed \citep{frederiks_ultraluminous_2013}. We convert these bolometric luminosities into spectral energies as per the method for SGRBs (with the LGRB width). We also assume all LGRBs to have a spectrum given by the Band energy function with indices $\alpha=-1$ and $\beta=-2.3$. 

\begin{figure}
    \centering
    \includegraphics[width=0.5\textwidth]{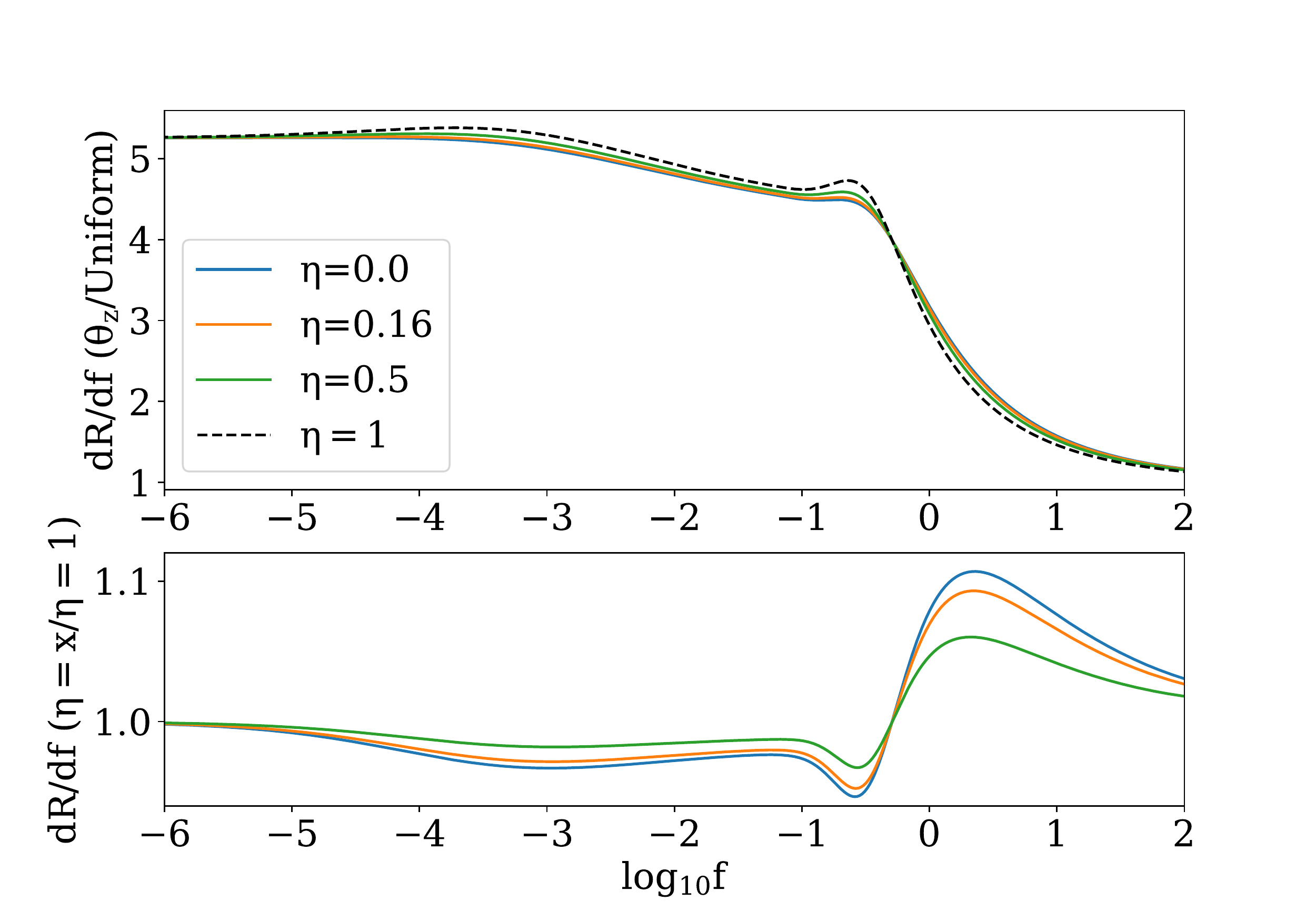}
    \caption{Same as Fig. \ref{fig:SGRB} but computed for LGRBs.}
    \label{fig:LGRB}
\end{figure}

Fig. \ref{fig:LGRB} shows $dR/df$ calculated for the LGRB event rate energy function described above. Similarly to the case of SGRBs, the absolute rate decreases towards higher fluences in line with the CSFR decreasing at lower redshifts, however the LGRB rate does not decrease as significantly towards lower fluences. The effect of lensing on the observed LGRB differential rate is similar to SGRBs at a maximum of $\approx 10\%$. The fractional change shows slightly more structure for the case of LGRBs due to the triple power law of $\theta_E$ but this does not substantially influence the effect from lensing, with any relative differences capped at a few percent.

\subsection{FRBs}\label{subsec:effectsFRBs}

For FRBs we assume that both $\theta_{E}$ and $\theta_\nu$ have single power law form described by indices $\gamma$ and $\alpha$ respectively. $\theta_E$ is also bounded by a hard cutoff at the maximum spectral energy $E_{\nu_e,\text{max}}$. Leading theories for FRB progenitors \citep{platts_living_2019} suggest that the central engine of FRBs is a compact stellar remnant such as a young magnetar. Such objects are also connected to massive star formation and hence, similarly to long GRBs, FRBs are expected to follow the CSFR. 

From \cite{james_fast_2021} the best fit values of these parameters are $\gamma=-2.16$, $\alpha=-1.5$ and $E_{\nu_e,\text{max}}=10^{32.84}$ erg/Hz. \cite{james_fast_2021} also allow for a redshift evolution on top of star formation by scaling the CSFR to the power of $n$. They find the best fit value of $n=1.77$ under the assumption that $\theta_\nu$ describes the change in energy of FRB bursts with frequency and not a change in the rate of bursts with frequency. Others within the field find differing model parameters. \cite{luo_frb_2020} assume a flat spectral distribution, and find $\gamma=1.79$, neglecting evolution (i.e. $n=1$) and \cite{shin_inferring_2022} find a shallower $\gamma=-1.3$. To capture these variations in behaviour our fiducial model will be $\gamma=-2.0$, $\alpha=-1.0$, $n=1.0$ and $E_{\nu_e,\text{max}}=10^{33}$ erg/Hz.

\begin{figure}
    \centering
    \includegraphics[width=0.5\textwidth]{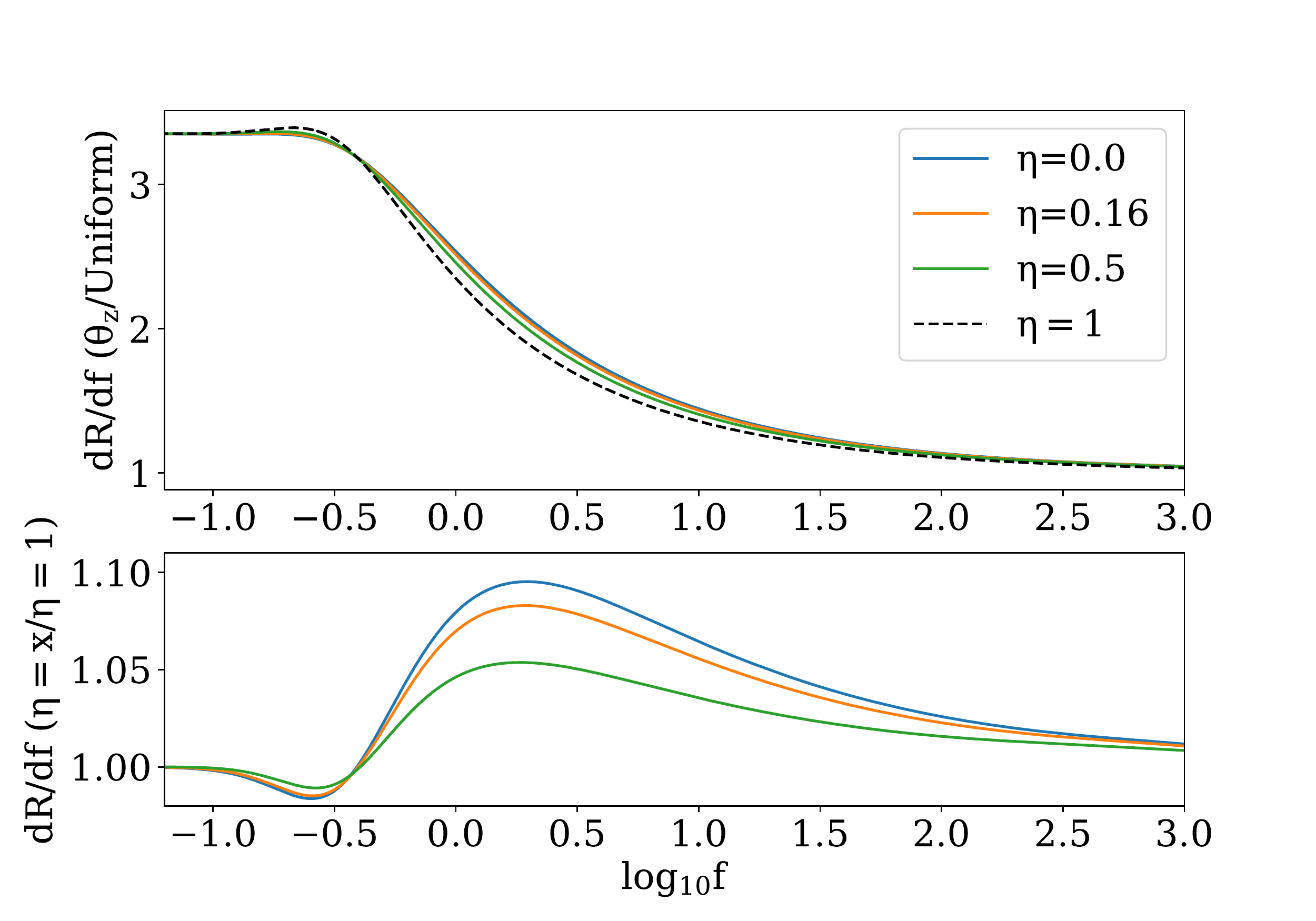}
    \caption{Sam as Fig. \ref{fig:SGRB} but computed for FRBs.}
    \label{fig:FRB}
\end{figure}

The $dR/df$ values resulting from the above calculations are depicted in Fig. \ref{fig:FRB}. The results show behaviour very similar to that of the GRB calculations, with a slightly higher minimum in the fractional change due to lensing. This similarity is unsurprising because we have assumed that both types of transient are related to star formation.

\section{Discussion}\label{sec:discussion}

Understanding the effects of gravitational lensing on $dR/df$ is crucial if the increasing number of recorded bursts with no redshift information are to be used to constrain $\Theta_E$. Most of the lensing effects we have derived here are small. In the current context of observational constraints on $\Theta_E$, for any of the transients mentioned here, the lensing effects are negligible compared to other sources of uncertainty. Hence, lensing may in most cases be ignored when calculating the expected differential rates from a $\Theta_E$ model.

Provided that there is a sharp\footnote{We take sharp to mean steeper than the power law dependence ($\xi$) of the magnification PDF with $\mu$. In the geometric limit $\xi\approx=-3$} cutoff in the intrinsic rate at some critical energy $E_{\nu_e,\text{max}}$, lensing will cause a fluctuation in $dR/df$ relative to what is expected in a smooth universe, effectively independent of the transients underlying $\Theta_E$. This fluctuation, seen in figures \ref{fig:SGRB}, \ref{fig:LGRB} and \ref{fig:FRB}, is a unique effect of lensing and could be used to constrain the value of $\eta$ which effects its scale. 

\begin{figure}
    \centering
    \includegraphics[width=0.5\textwidth]{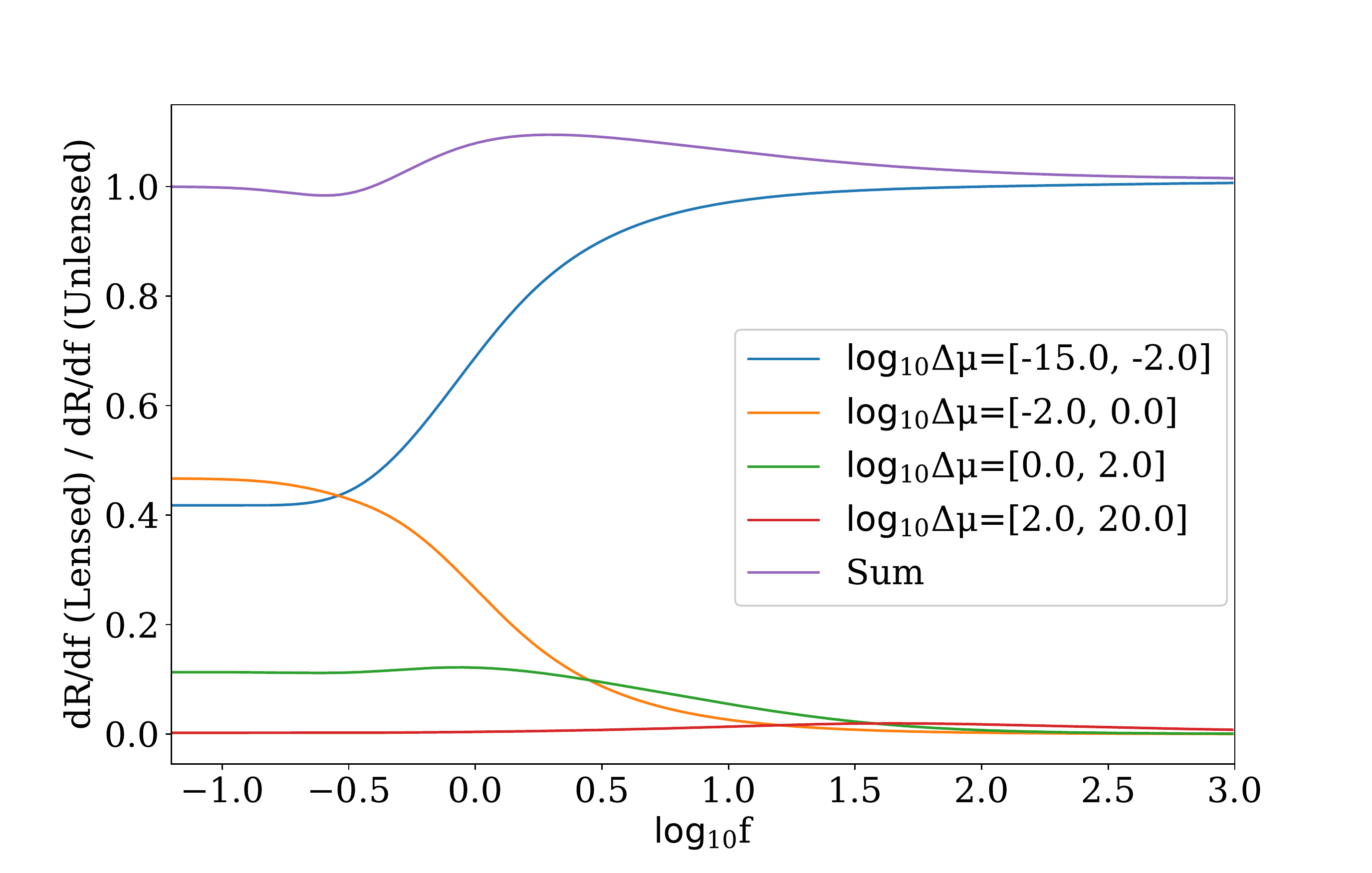}
    \caption{Fractional change in $dR/df$ due to lensing, as plotted in Fig. \ref{fig:FRB} decomposed into components of varying magnification. Upper and lower bounds correspond to the limits of our numerical method as described in \S \ref{subsec:numImplementation}.}
    \label{fig:FRBMagDecomp}
\end{figure}

A comparison between the figures in \S \ref{sec:IntrinsicParamVar} shows that the scale of the fluctuation is also dependent on $\theta_z$, $\alpha$ and $\gamma$. Due to this degeneracy, the intrinsic population parameters must be well known if $\eta$ is to be constrained from the observed rates. Given the strong dependence of the absolute rates on these parameters however, they will require far fewer transients to be well constrained. To avoid lensing effects when constraining the intrinsic parameters of $\Theta_E$, they should be modelled from low fluence bursts where the effect of lensing is negligible. To define a low fluence we require $f_{b,\eta}$, the break fluence defined in \S \ref{sec:fractionalChange}. Without knowing $E_{\nu_e,\text{max}}$ this becomes more difficult, however we can approximate a minimum value of $E_{\nu_e,\text{max}}$ and $f_b$ by considering the magnification decomposition of the fractional change in $dR/df$ due to lensing. For the FRB case shown in Fig. \ref{fig:FRB} this magnification decomposition is plotted in Fig. \ref{fig:FRBMagDecomp}. It shows that across all fluences there is very little contribution from magnifications above $\mu=10^2$. A lack of high magnification bursts means that $E_{\nu_e,\text{max}}$ is unlikely to be lower than $1/10^2$ the apparent maximum. By establishing a lower limit on $E_{\nu_e,\text{max}}$ and setting $f_{b,\eta}$ to correspond to this approximate maximum at a redshift of negligible star formation in a universe $\eta=0$ we can safely assume $f<f_{b,\eta}$ to be in the low fluence regime.

\begin{figure}
    \centering
    \includegraphics[width=0.5\textwidth]{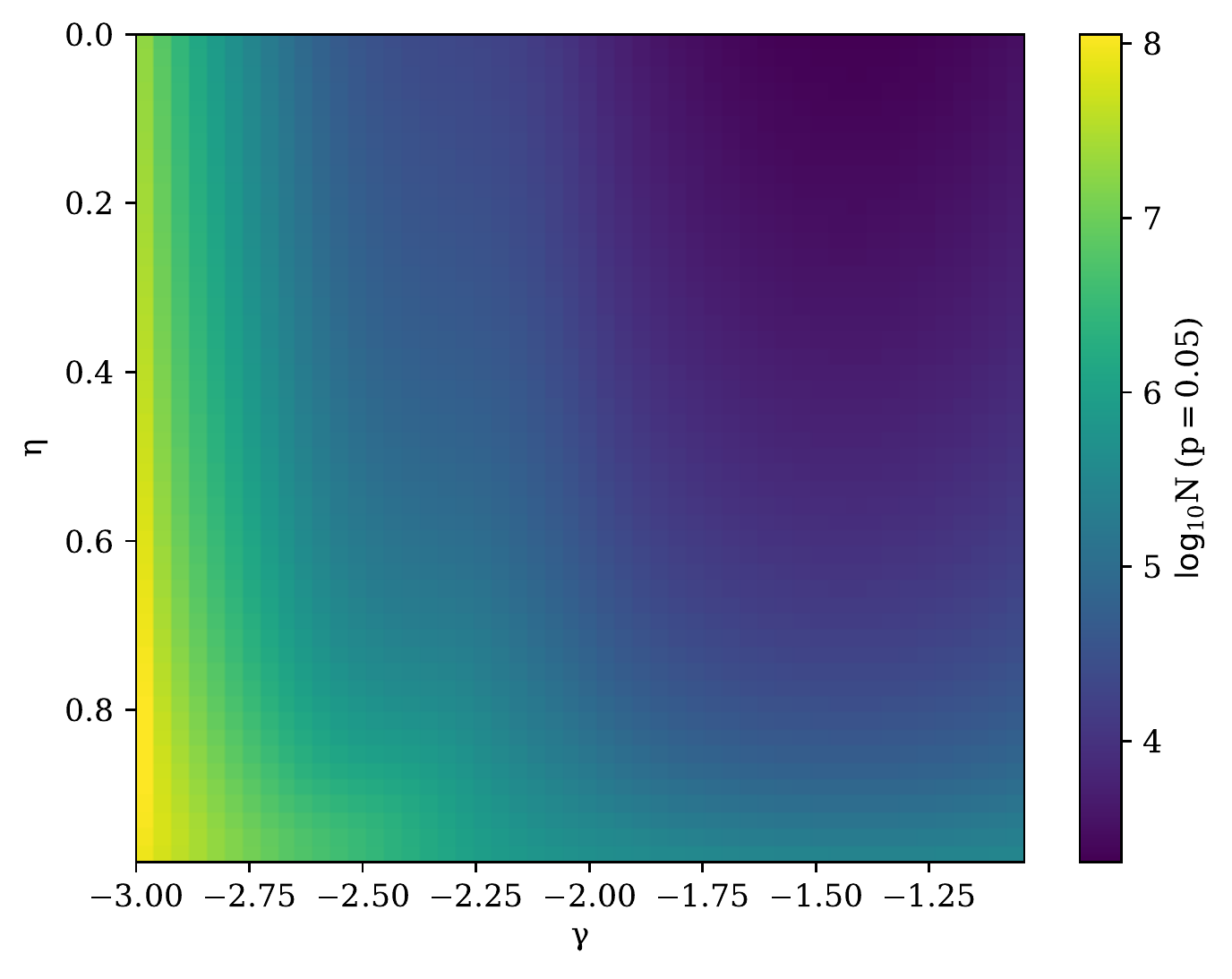}
    \includegraphics[width=0.5\textwidth]{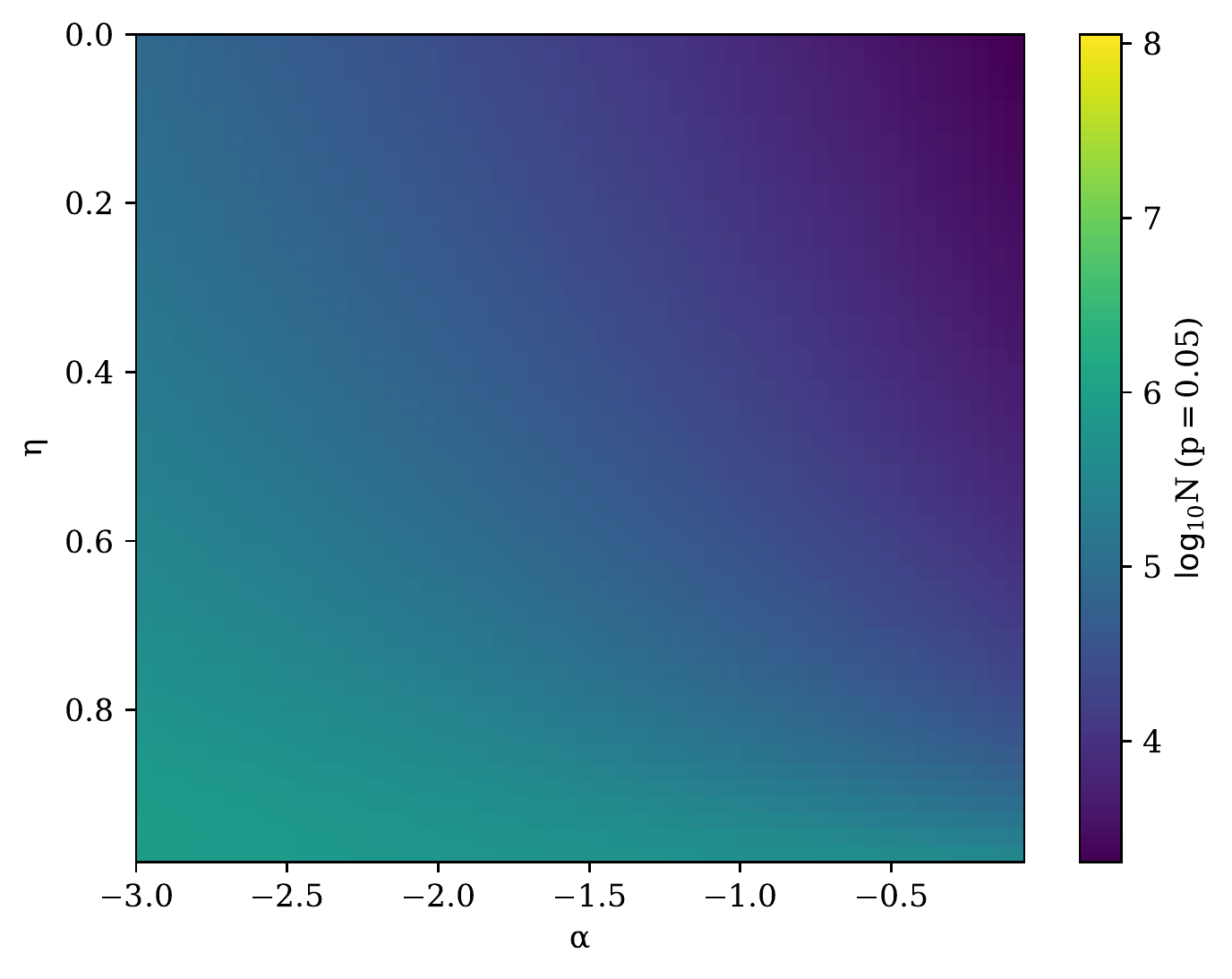}
    \caption{Number of high fluence FRBs required to distinguish the a universe with a smooth matter fraction $\eta$ from the $\eta=1$ case with $95\%$ confidence, assuming the intrinsic population parameters $\theta_z$, $\alpha$ and $\gamma$ are known. \textit{Top:} Plotted for varying values of $\gamma$ with an $\alpha=-1.0$ and $\theta_z\propto$ CSFR. \textit{Bottom:} Plotted for varying values of $\alpha$ with an $\gamma=-2.0$ and $\theta_z\propto$ CSFR.}
    \label{fig:numNeeded}
\end{figure}

Assuming that the population parameters are well constrained by these low fluence bursts, the intrinsic model could be extrapolated to the high fluence regime for the case of a smooth universe and compared to the observed differential event rates to place a lower limit of the value of $\eta$. Averaging over the expected fluctuation for all fluences higher than $f_{b,\eta}$, and below the fluence for a maximum energy burst at $z=0.001$, we can determine the number of high fluence bursts that would be required to statistically distinguish the expected average fluctuation. We plot this number for FRBs for varying values of $\eta$ and varying intrinsic $\gamma$ and $\alpha$ values in Fig. \ref{fig:numNeeded}. We assume that the observed bursts are distributed log-uniformly and calculate the relative error on the observed number as $1/\sqrt{N}$. We then calculate the number that would be required for the average of the absolute value of the fluctuation to be significant at the $95\%$ confidence level for a normal distribution (given the large number of bursts required we expect normality in the uncertainty). 

 As expected from Fig. \ref{fig:GenericFractional} and Fig. \ref{fig:GenericFractionalSpectral} the number of bursts required to distinguish a universe with a smooth matter fraction $\eta$ from a completely homogeneous universe $\eta=1$ generally increases with decreasing $\gamma$ and $\alpha$ with smoother universes naturally requiring more bursts. For the fiducial FRB population of $\gamma=-2.0$, and $\alpha=-1.0$ a universe comprised entirely of lenses can be ruled out using 8000 high fluence FRBs. Conversely a nearly smooth universe with 5$\%$ of its matter in lenses would require some $3.5\times10^5$ FRBs to distinguish from the smooth case. 
 
\begin{figure*}
    \centering
    \includegraphics{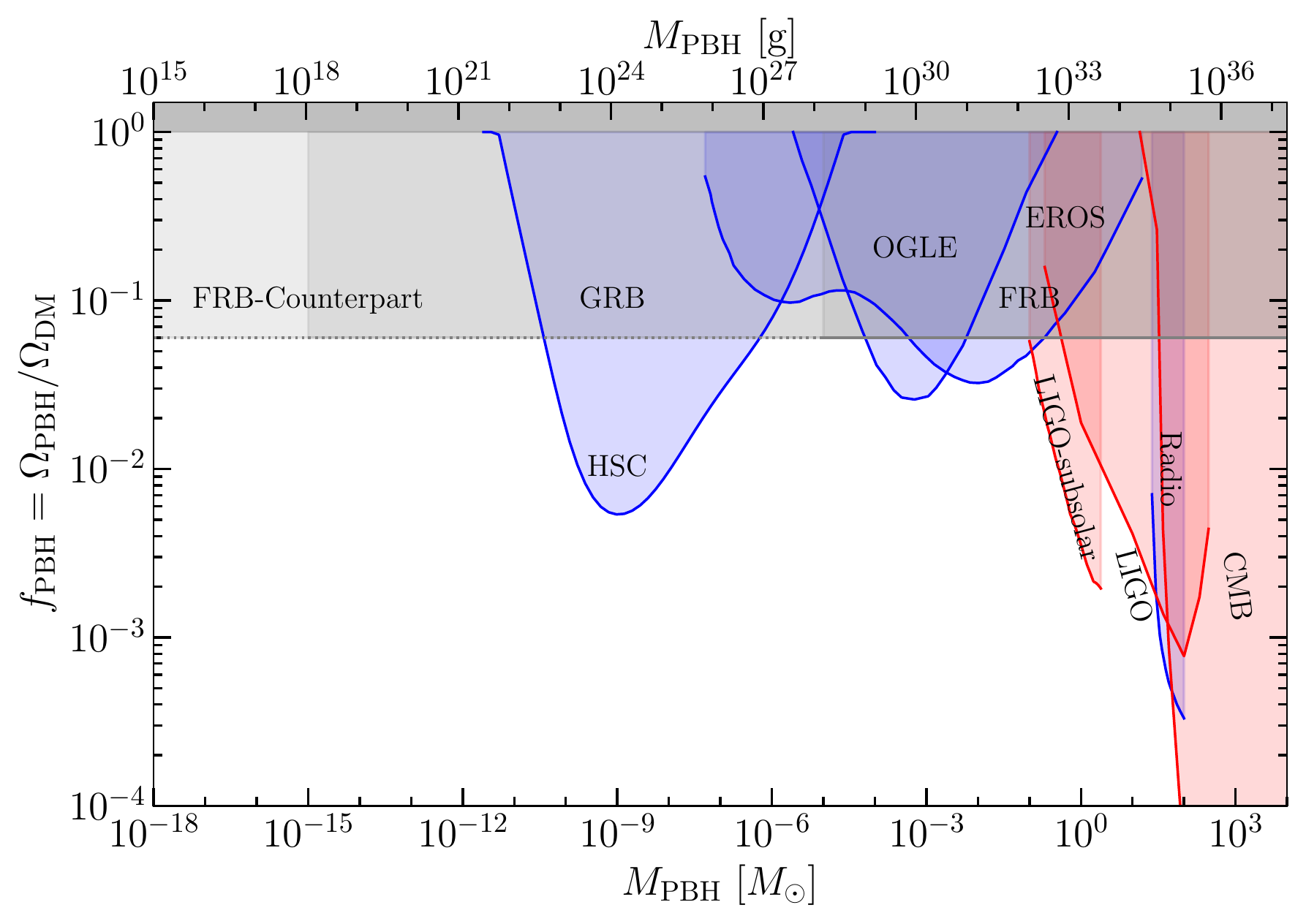}
    \caption{Current and potential constraints on the allowed fraction of dark matter in PBHs with a monochromatic mass function centred on $M_{PBH}$. Constraints based of measurements from our local galactic environment are shaded blue. Constraints that are cosmological in origin are shaded red. Our proposed constraints from $3.5\times10^5$ high fluence transients are shown in grey. These extend to varying lower mass limits based on observed frequency and source extent as detailed in appendix \ref{app:probedLensMass}. We note that $\eta=0.95$ corresponds to a maximum $f_{PBH}$ of $\approx6\%$ (for a Planck cosmology). This plot was made using code from the github repo \url{https://github.com/bradkav/PBHbounds} \citep[][and references therein]{kavanagh_bradkavpbhbounds_2019} }
    \label{fig:PBHBounds}
\end{figure*}

 Planned instruments such as CHORD or the proposed coherent all sky monitor (CASM) BURSTT \citep{lin_burstt_2022} have predicted detection rates of $\sim10^4$ per year \citep{connor_stellar_2022}. Several such instruments observing over the course of ten years could reasonably achieve our desired $3.5\times10^5$ high fluence FRBs, especially given the low-sensitivity -- high field of view mode of operation for CASMs. This would allow formation of broad and stringent constraints over parts of the PBH space that have only been probed locally. To show which masses the constraints apply over we must consider both source extension and wave optics effects as detailed in appendix \ref{app:probedLensMass}. Doing so we calculate the PBH dark matter fraction constraints shown in Fig. \ref{fig:PBHBounds} for $3.5\times 10^5$ high fluence FRBs observed at 1.4GHz. We highlight that the only observables required for each of these FRBs are the booleans $f>f_{b,\eta}$ and $\nu>\nu_{\text{min}}$\footnote{where this minimum frequency is used the establish the minimum probed lens mass}; a precise fluence measurement is not required, neither is a localisation or redshift. If these FRBs are observed at higher frequencies, these constraints will extend down to lower masses, with an infinite frequency FRB counterpart extending all the way down to $10^{-22}M_\odot$. Assuming a FRB-like functional form for the GRB intrinsic $\Theta_E$, a similar number of GRBs could constrain PBHs down to $10^{-15}M_\odot$ as displayed in the figure.

Constraining $E_{\nu_e,\text{max}}$ is an area of particular import in an inhomogeneous universe as lensing can have a large effect on the apparent maximum energy of a burst. $E_{\nu_e, \text{max}}$ is often taken to be the greatest apparent energy amongst observed bursts, which in an inhomogeneous universe can be a large overestimation. The compact nature of fast transient sources means they are susceptible to lensing by low mass objects. Such objects may leave no observational trace of the magnification they are causing. This makes it impossible to know the magnification of an individual burst and thus impossible to confidently approximate the intrinsic $E_{\nu_e,\text{max}}$ based on the largest apparent energy. Variation of $E_{\nu_e, \text{max}}$ however, is not degenerate with $\eta$ and hence may be constrained if sufficient data are collected to distinguish a fluctuation in $dR/df$ due to lensing.

\section{Conclusion}\label{sec:conc}
Gravitational lensing is one possible propagation effect to consider when modelling the differential events rate of fast transients from their intrinsic population functions. In doing so we have shown that, for a stationary universe:
\begin{enumerate}
    \item Except for the mass-energy range with prominent fringes shown in Fig. \ref{fig:AltScenCond}, FRBs are not all intrinsically low luminosity events highly magnified by gravitational lensing from point masses.
    \item Given current observational uncertainties, intrinsic population function parameters (other than $E_{\nu,\text{max}}$) inferred from observations without accounting for lensing will not significantly differ in a completely inhomogeneous universe ($\eta=0$)
    \item Wave optics may cause magnification PDFs to differ from the familiar $\mu^{-3}$ behaviour at high magnifications.
    \item For masses above 0.01 solar masses geometric optics will suffice for modelling the $dR/df$ of FRBs in an $\eta=0$ universe.
    \item Using low fluence ($<f_{b,\eta}$) observations of $dR/df$ to estimate $\Theta_E$ will be free from the effects of lensing. A further comparison with high fluence observations can be used to extract the influence of gravitational lensing. In this way the compactness of FRBs and GRBs can be exploited to constrain unexplored regions of dark matter parameter space such as low mass primordial black holes. We expect that 8000 high fluence, unlocalised FRBs would be required to rule out a completely clumpy universe, with $3.5\times10^5$ required to exclude more than $6\%$ of dark matter being in PBHs in the relevant mass range.
\end{enumerate}

\section*{Acknowledgements}
We thank Geraint F. Lewis and Ron Ekers for productive discussions on lensing. CMT is supported by an Australian Research Council Future Fellowship under project grant FT180100321. CWJ acknowledges support from the Australian Government through the Australian Research Council's Discovery Projects funding scheme  (project DP200102545).

\section*{Data Availability}

The data underlying this article are available in the article and in its online supplementary material.



\bibliographystyle{mnras}
\bibliography{references}



\appendix

\section{$D_{\eta}$}\label{app:Deta}
There are many models available for calculating cosmological distances in an inhomogeneous universe. We have opted here to use the ZKDR distance equation (Zel'dovich, Kantowski, Dyer/Dashevskii, Roeder, also known as the Dyer-Roeder distance), principally for its simplicity as an effective model rather than a full space-time description. More complicated models such as Swiss-cheese space-times are computationally limited to treating galaxy scale inhomogeneities \citep{fleury_interpretation_2013} and so are inadequate for addressing the scales we wish to consider here. Despite questions of the ZKDR distance’s validity \citep[e.g.][]{clarkson_misinterpreting_2012} the model has been shown analytically to be consistent with certain Swiss-cheese models \citep{fleury_swiss-cheese_2014} which are exact solutions to Einstein's field equations and achieves good agreement with more generalised models such as \cite{holz_new_1998} and \cite{bergstrom_lensing_2000}. 

The ZKDR distance equation calculates the value of $D_A$ for propagation through a void in an inhomogeneous, but on-average Friedmann-Lemaitre dust universe (dust refers to cold/non-relativistic matter) which we refer to as $D_\eta$. The equation, its derivation and the relevant boundary conditions can be found in \cite{kayser_general_1997}. The authors also present a general numerical method to solve for the ZKDR angular diameter distance ($D_\eta$) in an arbitrary cosmology. Their treatment makes three key assumptions. 
\begin{enumerate}
    \item The distribution of matter in the Universe can be divided into clumpy (inhomogeneous) and smooth categories, described by $\eta$, the fraction of the mass which is smooth.
    \item The beam subtended by the source contains no clumps.
    \item The light propagates far from all clumps, i.e. there is vanishing shear on the beam.
\end{enumerate}
Following their method we provide a simple numerical implementation for calculating the $D_\eta$ in the Python programming language \citep{rossum_python_2011}. $D_\eta$ can be solved for by considering the following system of coupled ordinary differential equations \citep{kayser_general_1997}:
\begin{align}
    D_{\eta}'(z)&=\frac{1}{(1+z)\sqrt{Q(z)}}\\
    D_{\eta}''(z)&=\frac{-\left(\frac{2Q(z)}{(1+z)+\frac{Q'(z)}{2}}\right)D'-\frac{3}{2}\eta\Omega_{M,0}(1+z)D}{Q(z)}\\
    Q(z)&=\Omega_{M,0}(1+z)^3-(\Omega_{M,0}+\Lambda_0-1)(1+z)^2+\Lambda_0\\
    Q'(z)&=3\Omega_{M,0}(1+z)^2-2(\Omega_{M,0}+\Lambda_0-1),
\end{align}
where primes denote derivatives with respect to redshift and $\Omega_{M,0}$ $\&$ $\Lambda_0$ are the matter density parameter and cosmological constant respectively at $z=0$. 

To solve the system we implement a numerical routine using SOLVE\_IVP \citep{virtanen_scipy_2020}. The source code for our implmentation can be found here :  \url{https://github.com/MWSammons/ZKDRDistance}. 

Also implemented in our function set is the generalised Dyer-Roeder model of \cite{linder_cosmological_1988}. This solution includes a treatment of relativistic matter and radiation densities in the universe. However, we note that results of this method achieve a worse agreement with the analytic solutions for a smooth universe in the case of a Planck cosmology compared to the forced-flat Kayser model.

\begin{figure}
\includegraphics[width=0.5\textwidth]{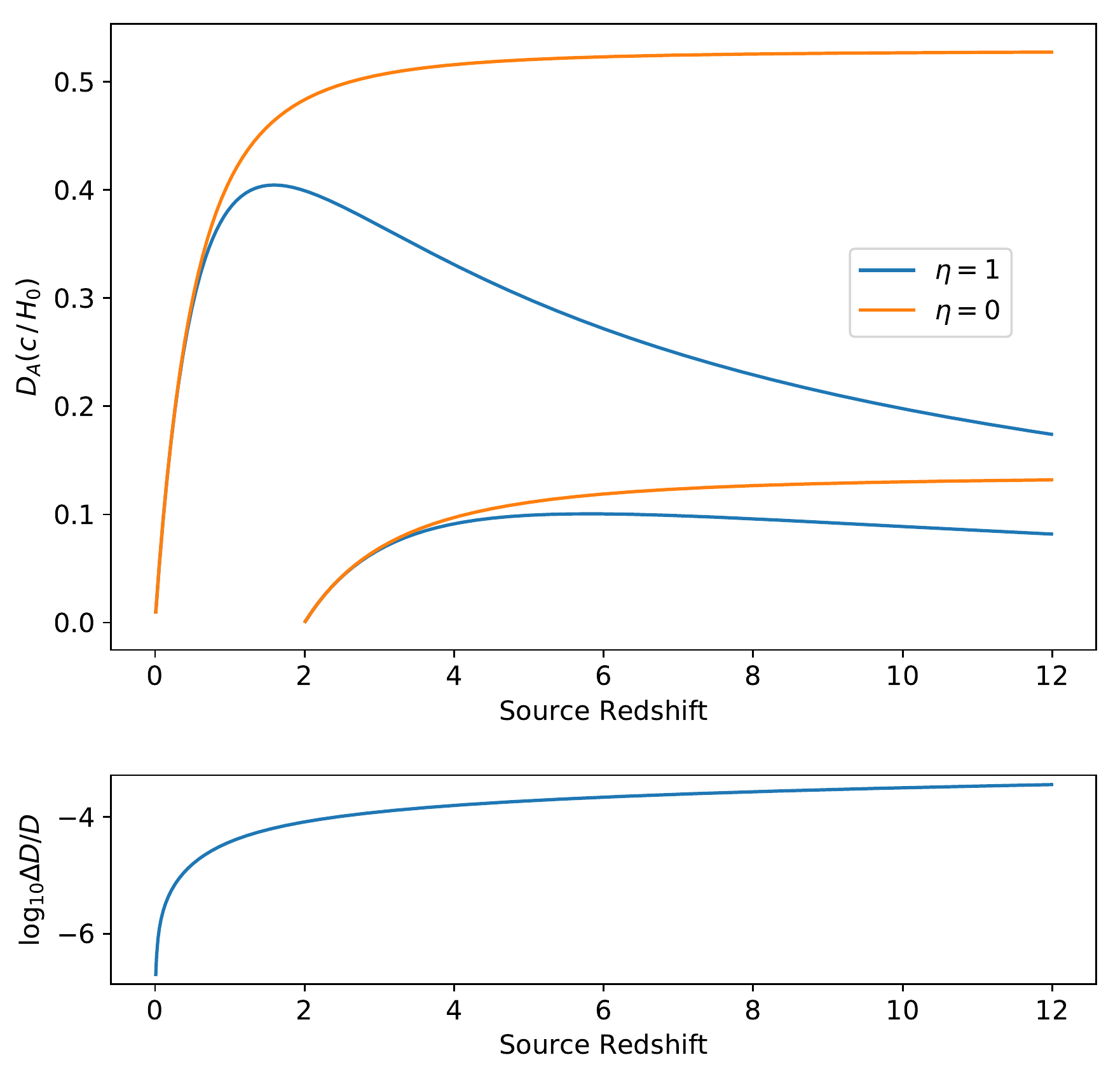}
\caption{(Top panel) Comparison of $D_1$ and $D_\eta$ for a completely inhomogeneous universe ($\eta=0$) in a Planck cosmology. Upper plots are for an observer at $z=0$, lower plots are for an observer at $z=2$. (Bottom panel) The log residuals between the analytic solution to $D_1$ \citep{hogg_distance_2000} and our numerical result for a Planck cosmology.}
 \label{fig:DistComparison}
\end{figure}

The results of our numerical approach are presented in Fig. \ref{fig:DistComparison}, which compares $D_\eta$ in a completely inhomogeneous universe ($\eta=0$) and $D_1$, for a Planck cosmology \citep{planck_collaboration_planck_2018}. For the remainder of this work $D_\eta$ will be calculated assuming this cosmology. To demonstrate the fidelity of our numeric method we also plot the residuals of $D_1$ with its analytic solution in the bottom panel of Fig. \ref{fig:DistComparison}. Our results show good agreement with the analytic solution. Moreover, we reproduce the large difference between $D_1$ and $D_\eta$ for $\eta=0$ at high redshift seen in Fig.~1 of \citet{kayser_general_1997}. One short-coming of the Kayser model is that any density not in cold matter $\Omega_M$ or dark energy $\Lambda$ implicitly contributes to the universe's curvature. Considering this, relativistic and radiation density in the Planck cosmology have been amalgamated into $\Omega_{M,0}$ to force a flat universe within the boundaries of the Kayser model.

\section{The Effect of Wave Optics}\label{app:WaveEffects}
Formally, geometric optics describes the behaviour of emission with an infinite frequency. In reality however, geometric optics provides an adequate description of gravitational lensing for all emission wavelengths much shorter than the gravitational radius of the lens. For the case of point mass lenses, \cite{oguri_strong_2019} defines a dimensionless parameter $w$ from this condition,
\begin{equation}\label{eq:dimlessFreq}
    w=2\pi f\frac{4GM(1+z_d)}{c^3},
\end{equation}
where $z_d$ is the redshift of the lens. 

The magnification for a point mass lens as a function of the source's angular impact parameter $\beta$ can then be defined for a wave optics regime as \footnote{We note this treatment is only valid when the geometric time delay along paths contributing to the interference pattern are much less than the pulses duration. Otherwise the boundary diffraction wave should be modelled explicitly \citep{born_principles_1999}}.
\begin{equation}\label{eq:waveAmplitude}
    \mu = \frac{\pi w}{1-e^{-\pi w}}\left|{}_{1}F_{1}\left(\frac{i}{2}w,1;\frac{i}{2}w \left(\frac{\beta}{\theta_E}\right)^2\right)\right|^2,
\end{equation}
where ${}_{1}F_{1}$ is the confluent hypergeometric function.

\begin{figure*}
    \centering
    \includegraphics[width=\textwidth]{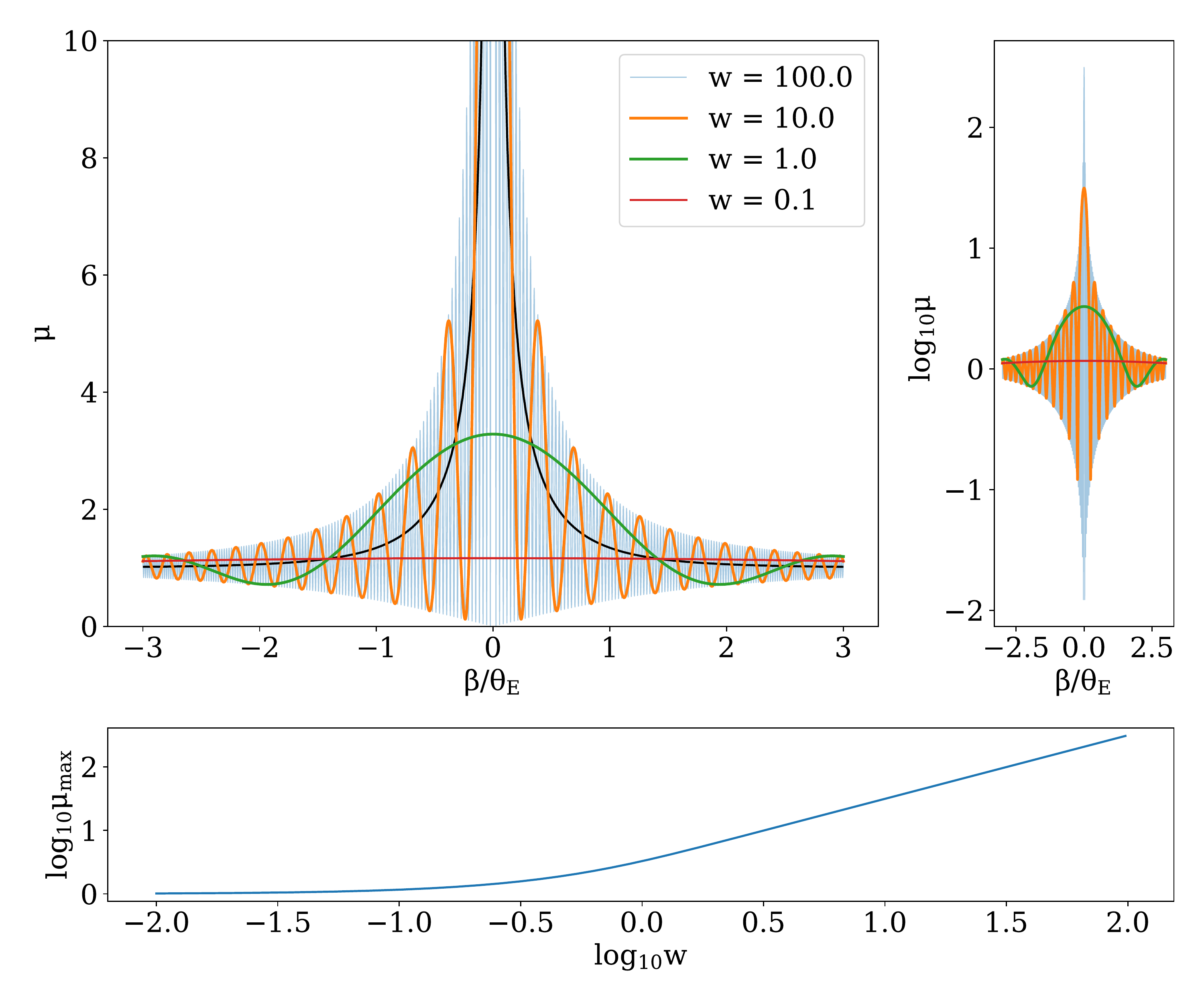}
    \caption{Behaviour of magnification in wave optics. \textit{(Top left/right)} magnification/log magnification of the source as a function of the source's angular impact parameter with respect to the optic axis centred on the lens, normalised by the einstein radius of the lens $\theta_E$, plotted for various dimensionless $w$ parameter choices. \textit{(Bottom)} Maximum magnification as a function of dimensionless parameter $w$, calculated using eq. (45) in \citep{oguri_strong_2019}.}
    \label{fig:waveAmplitude}
\end{figure*}

Fig. \ref{fig:waveAmplitude} shows this for a variety of $w$ values as well as the geometric case.   
For a $w$ value approaching infinity the wave and geometric optics results will agree, with a finite extent source smoothing over the infinitely compressed oscillations. As $w$ decreases so does the maximum possible magnification. Additionally, fringe spacing increases allowing sources of greater extent to exhibit an oscillatory magnification with $\beta$. 

If the optical depth to lensing is low, the magnification cumulative probability distribution function (CDF; $P(>\mu)$) will be proportional to the cross section of normalised angular impact ($\beta/\theta_E$) greater than a magnification $\mu$ \citep{turner_statistics_1984}. In the case of geometric optics, evaluating the cross section becomes simply
\begin{equation}\label{eq:geomCrossSec}
    \sigma=\pi \left(\frac{\beta_\mu}{\theta_E}\right)^2
\end{equation}
where $\beta_\mu$ is the angular impact parameter of the source at a magnification $\mu$. $\beta_\mu$ can be determined from the black line on Fig. \ref{fig:waveAmplitude}, which is governed by eq. (2.5) in \cite{turner_statistics_1984}. Ultimately this yields the conventional $dP/d\mu\equiv p(\mu)\propto \mu^{-3}$ behaviour.

\begin{figure}
    \centering
    \includegraphics[width=0.5\textwidth]{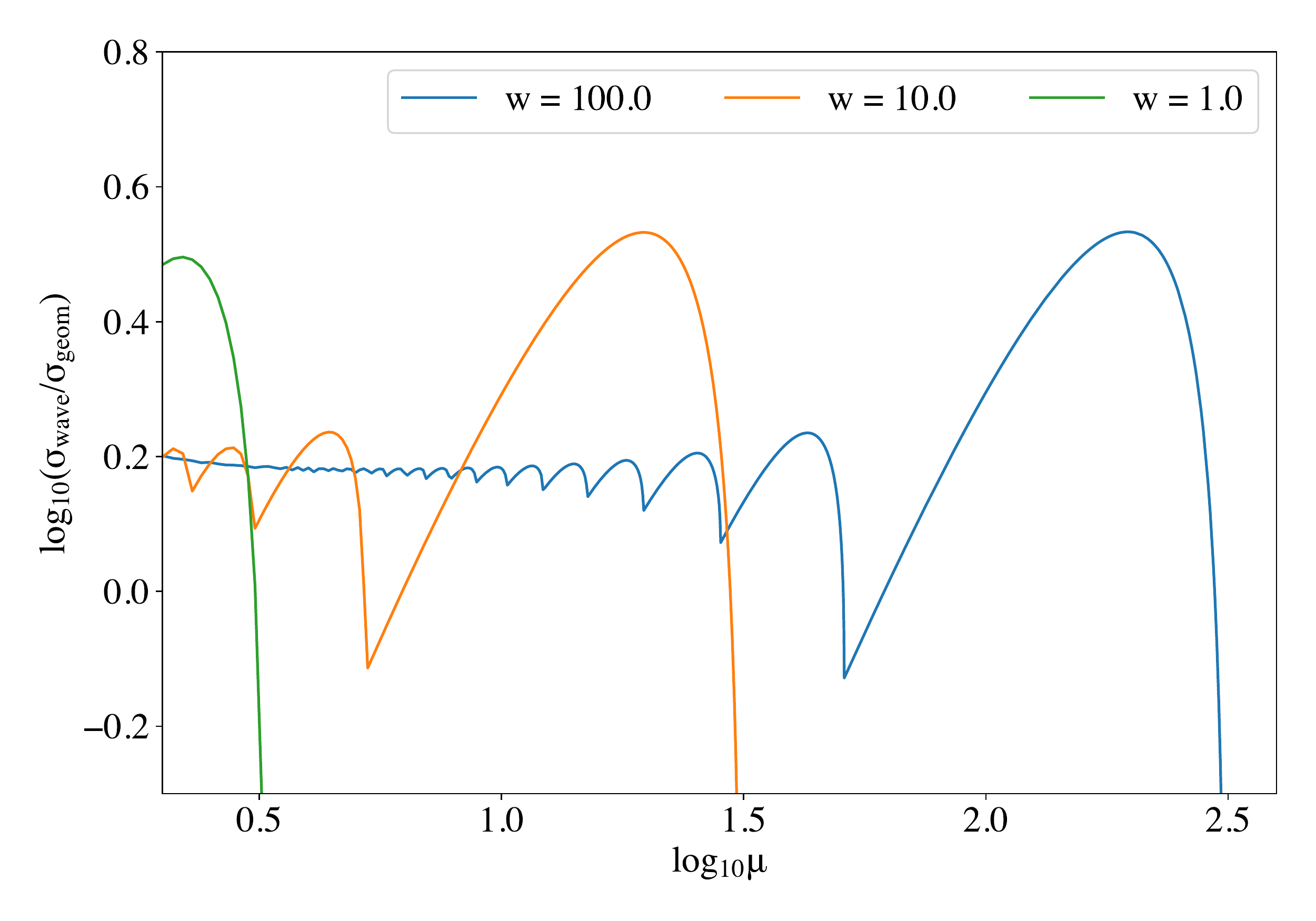}
    \caption{Cross section to lensing above a magnification $\mu$ calculated from the wave optics magnifications shown in Fig. \ref{fig:waveAmplitude}, normalised by the corresponding cross section derived using geometric optics (eq. (\ref{eq:geomCrossSec})). $w$ values are calculated from eq. (\ref{eq:dimlessFreq}). Cross sections are directly proportional to the magnification CDF.}
    \label{fig:waveCDFNorm}
\end{figure}

As can be intuited from Fig. \ref{fig:waveAmplitude}, the cross section to lensing above $\mu$ for lower values of $w$ is significantly more complicated. By integrating the ring element $2\pi\beta d\beta$ for all magnifications above $\mu$ we can determine the cross section corresponding to $P(>\mu)$ for any $w$. We plot this cross section in Fig. \ref{fig:waveCDFNorm}, normalised by its geometric counterpart to provide a comparison between the results of geometric and wave optics.

Fig. \ref{fig:waveCDFNorm} shows that the normalised cross section over the range of $\mu$ and $w$ values plotted is often significantly above 1. This means that $P(>\mu)$ is often greater in the wave optics regime than the geometric regime. This is in agreement with wave optics results derived by \cite{jow_wave_2020}.

For high values of $w$, the low magnifications behave very similarly to the geometric case (i.e. the normalised cross section is flat with $\mu$). As the magnification increases however, the scale of cross section oscillation increases and subsections of the magnification space begin to deviate significantly from the flat geometric behaviour until the maximum magnification is reached. Decreasing $w$, either by decreasing the emission frequency or decreasing lens mass then effectively translates the normalised cross section to lower magnification. 

If the frequency of emission or lens mass is low enough such that regions of significant oscillation in normalised cross section are present at observed magnifications, then geometric optics should not be applied to calculate the expected effects of lensing. For an FRB emitted at 1\;GHz, at redshift 0.1, lensed by a 0.01 $M_{\odot}$ point mass, $w\sim10^3$ and the maximum magnification will be $\approx10^{3.5}$. From Fig. \ref{fig:waveCDFNorm} we can see that for the $w=100$ case, magnifications a factor of $10^{1.5}$ below the maximum (at around log$_{10}\mu=1.0$) have geometric like behaviour. Applying this same condition to our canonical FRB case, it is reasonable to assume that magnifications below $\sim10^{2}(10^{3.5}/10^{1.5})$ will have $p(\mu)\propto \mu^{-3}$. Taking the results of our lensed $dR/df$ calculation shown in Fig. \ref{fig:FRB} and decomposing it into its components in magnification space we get Fig. \ref{fig:FRBMagDecomp}. From this figure we can see that the fractional change in $dR/df$ due to lensing is dominated by low magnifications. Specifically, more than $98\%$ of the total comes from magnifications less than 100 at all fluences. This suggests that the fraction of observed sources at magnifications above $10^{2}$ will be negligible and therefore that our results for FRB lensing, using the geometric $p(\mu)\propto\mu^{-3}$ should apply for lens masses greater than 0.01$M_\odot$\footnote{We have not accounted for the increased cross section size in the case of a wave optics and so our results will underestimate the effect of lensing}. 

If observed bursts are dominated by lensing at magnifications where the cross section to lensing shows prominent fringes, e.g. $\mu=[10-300]$ for $w=100$, the true lensing PDF could have behaviour significantly different from $p(\mu)\propto \mu^{-3}$. In this context, counter to the discussion in \S \ref{sec:AllLensed}, all observed bursts could be highly magnified, despite an observed $\gamma\neq-3$. The associated decrease in $\mu_{\text{max}}$ however restricts the parameter space where this could occur. Using $w=100$, a potential example could be FRBs at an emission frequency of 1\;GHz requiring a magnification above $\sim10$ but below $\sim300$ to be observed in a universe populated by $\sim 10^{-3}M_\odot$ mass lenses ($w\approx 100$). In such a scenario all observed FRBs would be lensed but the energy index $\gamma$ could differ from the expected $-3$ value. These FRBs would also only be observable above $\sim30$ MHz, at which point $\mu_{\text{max}}\approx10$.

\section{Probed Lens Masses}\label{app:probedLensMass}
Present constraints indicate that the matter distribution must be smooth when averaged over volumes comparable to or larger than the beam defined by a SN Ia near maximum light ($1000$ pc$^3$). In treating inhomogeneity below this smoothing scale it is instructive to consider a field of homogeneously distributed clumps of mass $M_c$ composing some fraction ($f$) of the Universe's total matter density $\Omega_{\text{c}}=f\Omega_{M,0}$. We can then characterise the level of inhomogeneity by comparing $M_c$ to $M_\text{beam}$, the mass enclosed by the beam. For the case of a smooth mass distribution ($\eta=1$) in a flat universe (k=0), 
\begin{equation}\label{eq:MCchar}
    M_{\text{beam}}=\rho_{cr,0}d_H^3\frac{A}{D_A^2(z_S)}\Bigg\{\Omega_{c,0}\int\limits_0^{z_S}\frac{(1+z)^2\tilde{D_A}^2(z)}{E(z)}dz\Bigg\},
\end{equation}
as per the comoving volume equation in \cite{hogg_distance_2000}. $\rho_{\text{cr},0}$ is the critical density at $z=0$, A is the area of the source, $d_H$ is the Hubble distance, $\tilde{D}$ denotes a distance normalised by $d_H$, $E(z)=H(z)/H_0$, and $z_S$ is the source redshift.

For a $M_c\ll M_{\text{beam}}$ the expected number of clumps within the beam will be $\langle N\rangle\gg 1$. As our distribution of clumps has constant co-moving density, the random fluctuations in $N$ will follow Poisson noise with standard deviation of $\sqrt{N}$, making the fractional fluctuation in both $N$ and the total convergence of the beam small for large $\langle N\rangle$. It is therefore unlikely, in the case of $M_c\ll M_{\text{beam}}$, to observe $D_A$ significantly different from $D_1$. As the value of $M_c$ increases, the fractional fluctuation in $N$ also increases, eventually yielding a significant probability of a beam containing no clumps. For the case of $M_c\gg M_{\text{beam}}$, a beam is most likely to contain no clumps in which case $D_\eta$ will apply.

\subsection{Magnification of Extended Sources}
The exception to our $M_\text{beam}$ criteria would be when many clumps lie within the beam and each causes a significant magnification of the source. This scenario may be observationally distinct from the smooth matter case and hence the requirement of $M_c\ll M_{\text{beam}}$ is a necessary but not sufficient condition for smoothness. In order to treat a matter distribution as though it were smooth we must also require that the maximum magnification by any lenses within its volume be low. Because the size of lensing masses ($M_c$) in question are exceedingly low it is appropriate to consider the finite size of even our most compact sources. 

Extended sources can be significantly magnified if their angular size ($\theta_S$) is comparable to the Einstein angle of the lens ($\theta_E$). As shown by \cite{schneider_gravitational_1992}, the maximum magnification ($\mu_{\text{max}}$) from an extended source is given by, 
\begin{equation}\label{eq:magMax}
    \mu_{\text{max}}=\frac{\sqrt{4+r^2}}{r},
\end{equation}
where $r=\theta_S/\theta_E$. If these two angles are equal ($r=1$) then $\mu_{\text{max}}\approx2.24$, dropping approximately linearly with $r$. Using this equivalence, we can determine the mass of clumps, below which only small magnifications will be observed,
\begin{equation}\label{eq:MCchar2}
    M_{\text{lens}} = \rho_{cr,0}d_H^3\frac{A}{D_s^2}\Bigg\{\frac{2}{3}\frac{\tilde{D_d}\tilde{D_s}}{\tilde{D_{ds}}}\Bigg\}.
\end{equation}

Uniform mass distributions with clump masses that are then below both $M_\text{beam}$ and $M_\text{lens}$ will have clumps both numerous within the beam and able to affect only a low maximum magnification of the source. Such distributions will be largely indistinguishable from a smooth matter distribution.

The linear density field associated with a uniform distribution of clumps has a vanishing shear due to matter outside the beam \citep{nakamura_effect_1997}. As such, for $M_c\gg\min[M_\text{lens}, M_\text{beam}]$ the assumptions of the ZKDR distance model are satisfied and we can calculate $D_A$ for a beam without clumps using the method of \cite{kayser_general_1997} (i.e. $D_A=D_\eta$, with $f=\eta$). We then expect a source averaged magnification with respect to the empty beam $\langle\mu\rangle=\frac{D_\eta^2}{D_1^2}$ and the most likely line of sight to a source to be characterised by $D_\eta$. 

\begin{figure*}
    \centering
    \includegraphics[width=0.9\textwidth]{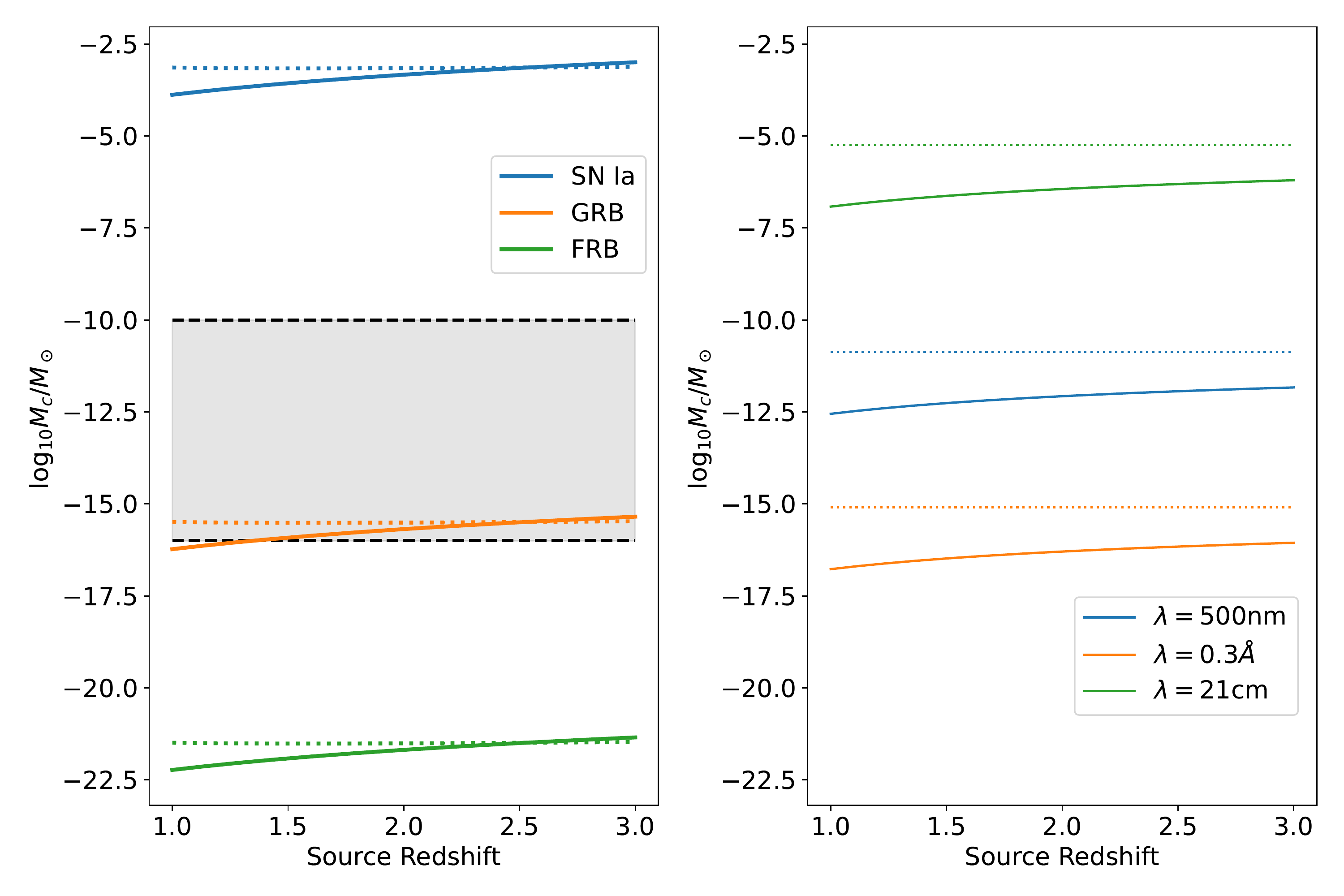}
    \caption{$M_\text{beam}$ and $M_{\text{lens}}$ limits to clump mass plotted with full and dotted lines respectively. For $M_c\gg\min[M_\text{beam},M_\text{lens}]$ the mass will appear inhomogeneous and $D_\eta$ will describe the most likely line of sight to a source. For $M_c\ll\min[M_\text{beam},M_\text{lens}]$ the mass will appear homogeneous and all lines of sight will be described by $D_1$. \textit{Left}: Constraints are calculated in the limit of geometric optics for source size representing SN Ia ($100$ AU), GRBs ($10^4$ km) and FRBs ($10$ km), using equations (\ref{eq:MCchar}) and (\ref{eq:MCchar2}). The shaded region gives the range of PBH masses that are observationally unconstrained \textit{Right}: Limits are calculated considering physical optics for wavelengths representing SN Ia, GRBs and FRBs, using equations (\ref{eq:diffMBeam}) and (\ref{eq:diffMLens}). A mass distribution may be considered smooth if $M_c\ll\min[M_\text{beam},M_\text{lens}]$ for either of the geometric or wave optics limits, hence wave optics will be the dominant limit for FRBs observed at radio frequencies.}
    \label{fig:ProbedRange}
\end{figure*}
In the left panel of Fig. \ref{fig:ProbedRange} we visualise the range of clump masses which should and should not be considered smooth by plotting $M_\text{beam}$ (full lines) and $M_\text{lens}$ (dotted lines) for various sources. We have used the canonical source sizes for SN Ia, GRBs and FRBs ($100$ AU, $10^4$ km and $10$ km, respectively) to calculate each criteria. Additionally, for the calculation of $M_{\text{lens}}$ we assume that $D_{ds}=D_d$ as this will capture the region of lens geometry with the highest contribution to the lensing optical depth \citep{turner_statistics_1984}. 

As expected from the similarity of equations (\ref{eq:MCchar}) and (\ref{eq:MCchar2}), the value of each criteria is relatively similar over the vast range of masses we are considering. For each source $M_\text{beam}$ is shown to be the dominating criteria over much of the redshift space of interest. Our interest is restricted to source redshifts $1.0<z_S<3.0$ as the difference between $D_0$ and $D_1$ for $z<1$ is small as per Fig. \ref{fig:DistComparison} and we expect few sources to be observed at higher redshifts.

We note that a caveat of this model is that as the convergence of the beam fluctuates with $N$, the apparent angular size of the source will also fluctuate. This leads to changes in both the beam's volume and consequently in $N$. We do not account for this second order effect, however qualitatively the resulting change to $N$ will be in the same direction as the original fluctuation. This will cause an increase to the standard deviation of the distribution of $N$ and therefore an increase in the level of inhomogeneity. By neglecting this second order effect our conclusions on the mass range of inhomogeneities each source is sensitive to will be conservative.

Any objects having masses in the stellar range constitute inhomogeneities for SN Ia, GRBs and FRBs, as can be seen from Fig \ref{fig:ProbedRange} (left), and for that reason a large population of such objects is already excluded by the SN Ia data \citep{helbig_m-z_2015}. Visible stars themselves amount to only a small fraction of the average matter density, $\Omega_{\text{stars}}/\Omega_{\text{Matter}}\sim0.01$ \citep{fukugita_cosmic_2004}, and a uniformly distributed population at this low level would not have a substantial effect on the angular diameter distances; this case would be well approximated by the $\eta=1$ calculations shown in Figure 3. In fact visible stars are far from uniformly distributed; they are concentrated in the central regions of galaxies and so can play a major role as gravitational lenses on some particular lines-of-sight, but have little influence on the background geometry.

The shaded region in Fig. \ref{fig:ProbedRange} (left) corresponds to the range of PBH masses which could theoretically still constitute 100 $\%$ of our Universe's dark matter. Notably, this region lies well below $M_\text{beam}$ for SN Ia, meaning that current observational constraints are insensitive to PBHs in the asteroid to sub-lunar range. Conversely, the region lies far above $M_\text{beam}$ for FRBs, with a majority being far above $M_\text{beam}$ for GRBs as well. Thus, if dark matter were comprised mostly of PBHs in the unconstrained range, $D_A$ for FRBs and GRBs would be affected. This suggests that compact cosmological transient such as FRBs and GRBs could provide a new way to constrain dark matter in this unexplored range.

\subsection{Diffraction Limitations}
As suggested by several authots \cite{oguri_strong_2019, jow_wave_2020} wave effects may also be important to the lensing of FRBs. In the context of our previous two constraints given by equations (\ref{eq:MCchar}) and (\ref{eq:MCchar2}), considering physical optics will have two effects: it will set a minimum probed volume corresponding to the Fresnel scale, and it will set a maximum amplification as described below.

Under wave optics, radiation from the source will sample a transverse area corresponding to the Fresnel scale. Hence, the volume probed by a source cannot be smaller than the Fresnel zone integrated over the line of sight. Using this volume we can recalculate $M_\text{beam}$ as
\begin{equation}\label{eq:diffMBeam}
    M_\text{beam}=\rho_{cr,0}\,\lambda d_H^2\Bigg\{\Omega_c\int \frac{\tilde{D_{ds}}\tilde{D_d}}{\tilde{D_s}}\frac{(1+z)^2}{E(z)}dz\Bigg\}
\end{equation}
where $\lambda$ is the wavelength of the radiation. 

Diffraction around a lens will also set the maximum amplification\footnote{We note that amplification here refers directly to wave amplitude rather than magnification which is defined with respect to the angular size of the image} we can observe from a lens' magnification. When the Schwarzschild radius is equivalent to the wavelength of the emitted radiation the maximum amplification will be $I_\text{max}\approx 3.28$ \citep{nakamura_effect_1997}. As we did earlier for extended sources we can use this as a fiducial point and calculate the mass below which diffraction will significantly restrict amplification,

\begin{equation}\label{eq:diffMLens}
    M_{\text{lens}}=\frac{c^2\lambda}{8\pi G}.
\end{equation}

Just as for our previous constraints, clump mass $M_c\ll \min[M_\text{beam}, M_\text{lens}]$ will be numerous within the Fresnel volume and have low maximum magnifications allowing their distribution to be effectively treated as smooth for the purpose of calculating distance measures. 

The right panel of Fig. \ref{fig:ProbedRange} shows the $M_{\text{beam}}$ (full lines) and $M_{\text{lens}}$ (dotted lines) criteria calculated for a range of representative wavelengths for the prompt emission from each of the sources in the left panel ($500$ nm = optical = SN Ia, $0.3$\AA = gamma-ray = GRB, $21$ cm = radio = FRB). Between the two limits we can see that in the redshift range of interest $M_\text{beam}\gg M_\text{lens}$. Consequently, for these redshifts it is sufficient to say that if the uniformly distributed clumps are numerous within the Fresnel volume then they may be treated as a smooth distribution of matter. Comparing the results between panels we can see that the mass limits calculated in the geometric optics limit dominate over their physical optics counterparts for both SN Ia and GRBs, i.e. small inhomogeneities will be smoothed over by the source sizes before wave effects become important. For FRBs however, diffraction will smooth over inhomogeneities far larger than what could be probed on the basis of their source size alone. This leaves only a narrow range of possible inhomogeneites they could probe that are not already ruled out from SN Ia observations.

However, despite their curtailed potential in the radio, the results obtained for FRBs in the geometric optics case further motivate multi-wavelength observations of FRBs. Observations of the so called Galactic FRB have shown coincident x-ray emission with the prompt radio burst \citep{ridnaia_peculiar_2021, the_chimefrb_collaboration_bright_2020}. Such a high frequency counterpart would drastically reduce the diffraction limit associated with FRB observations at $21$ cm, allowing FRBs to probe a similar range of inhomogeneities as GRBs.

\section{Derivations}\label{app:derivations}
\subsection{Differential Rates in a Smooth Universe}\label{app:derivations:SmoothDiffRate}
A small observed rate can be expressed using the event rate energy function of the fast transient population $\Theta_E$ as,
\begin{align}
    dR &= \Theta_E(E_{\nu_e}, z, \nu_e)dE\;dV_c
    \intertext{the comoving volume element is given by}
    V_c &= \frac{4}{3}\pi D_c^3(z)
    \intertext{differentiating with respect to redshift gives}
    \frac{dV_c}{dz}&=4\pi D_c^2(z)\frac{dD_c}{dz}
    \intertext{Intrinsic spectral energy ($E_{\nu_e}$) is given by}
    E_{\nu_e} &= 4\pi D_L^2(z)\frac{F_\nu}{(1+z)^2} 
    \intertext{where $F_\nu$ is the observed fluence of the transient at observation frequency $\nu$, the factor of $1/(1+z)^2$ accounts for bandwidth compression by cosmological redshift as well as the dilation of the bursts duration in time, differentiating with respect to observed fluence and holding redshift constant gives us}
    \frac{\partial E_{\nu_e}}{\partial F_{\nu}} &= 4\pi D_L^2(z)\frac{1}{(1+z)^2}
    \intertext{putting these components into a integration over redshift transforms our partial differential equation into a full differential equation, yielding}
    \frac{dR}{dF_\nu}&=\int dz\; 16\pi^2D_L^2(z)D_c^2(z)\frac{1}{(1+z)^3}\frac{dD_c}{dz}\Theta_E(E_{\nu_e}, z, \nu_e)
    \intertext{where $\nu_e$ can be expressed as $\nu_e=(1+z)\nu$, sampling from the emission frequency region of the energy function as opposed to the observation frequency implicitly handles the required k-correction. The additional factor of $1/(1+z)$ accounts for the redshift of the burst rate itself.}
    \frac{dR}{dF_\nu}&=\int dz\; 16\pi^2D_L^2(z)D_c^2(z)\frac{1}{(1+z)^3}\frac{dD_c}{dz}\Theta_E(E_{\nu_e}, z, (1+z)\nu)
\end{align}

\subsection{Differential Rates in a Clumpy Universe}\label{app:derivations:ClumpyDiffRate}
Similarly to above, a small observed rate of fast trasnients can be expressed for an inhomogeneous universe using the event rate energy function of the transient population and the probability of magnification by a factor $\mu$ from gravitational lensing as,
\begin{align}
    dR &= p(\mu, z)\Theta_E(E_{\nu_e}, z, \nu_e)dE\;dV_c
    \intertext{the comoving volume element is given by}
    V_c &= \frac{4}{3}\pi D_c^3(z)
    \intertext{differentiating with respect to redshift gives}
    \frac{dV_c}{dz}&=4\pi D_c^2(z)\frac{dD_c}{dz}
    \intertext{Intrinsic spectral energy ($E_{\nu_e}$) is given by}
    E_{\nu_e} &= 4\pi (D_\eta(1+z)^2)^2\frac{F_\nu}{\mu(1+z)^2} 
    \intertext{where $F_\nu$ is the observed fluence of the transient at observation frequency $\nu$, $D_\eta(1+z)^2$ is the luminosity distance in an inhomogeneous universe with a smooth matter fraction $\eta$ and the factor of $1/(1+z)^2$ accounts for bandwidth compression by cosmological redshift as well as the dilation of the bursts duration in time. Differentiating with respect to observed fluence and holding redshift constant gives us}
    \frac{\partial E_{\nu_e}}{\partial F_{\nu}} &= 4\pi (D_\eta(1+z)^2)^2\frac{1}{\mu(1+z)^2}
    \intertext{putting these components into a integration over redshift transforms our partial differential equation into a full differential equation, yielding}
    \frac{dR}{dF_\nu}&=\int\;dz 16\pi^2(D_\eta(1+z)^2)^2D_c^2(z)\frac{1}{(1+z)^3}\frac{dD_c}{dz}\\
    \times &\int\;d\mu\;\frac{1}{\mu}p(\mu,z)\Theta_E(E_{\nu_e}, z, \nu_e)
    \intertext{where $\nu_e$ can be expressed as $\nu_e=(1+z)\nu$, sampling from the emission frequency region of the energy function as opposed to the observation frequency implicitly handles the required k-correction. The additional factor of $1/(1+z)$ accounts for the redshift of the burst rate itself.}
    \frac{dR}{dF_\nu}&=\int\;dz 16\pi^2(D_\eta(1+z)^2)^2)D_c^2(z)\frac{1}{(1+z)^3}\frac{dD_c}{dz}\\
    &\times\int\;d\mu\;\frac{1}{\mu}p(\mu,z)\Theta_E(E_{\nu_e}, z, (1+z)\nu)
\end{align}


\bsp	
\label{lastpage}
\end{document}